\documentclass[prd,twocolumn,showpacs,nofootinbib]{revtex4}
\usepackage{natbib,amsmath,graphicx,bm,color}
\def\mnras{Mon. Not. R. Astron. Soc.}

\def\apjs{Astrophys. J. Suppl.}
\def\physrep{Phys. Rep.}

\def\aap{Astron. Astrophys.}

\begin{document}
\title{
  Massive Neutrinos in Cosmology: Analytic Solutions and Fluid Approximation
}
\author{Masatoshi Shoji \& Eiichiro Komatsu}
\affiliation{Texas Cosmology Center and Department of Astronomy, University of Texas at Austin, \\
  1 University Station, C1400, Austin, TX, 78712}
\date{\today}
\email[]{mshoji@astro.as.utexas.edu}
%%%%%%%%%%%%%%%%%%%%%%%%%%%%%%%%%%%%%%%%%%%%%%%%%%%%%%%%%%%%%%%%%%
\begin{abstract}
We study the evolution of linear density fluctuations of free-streaming massive neutrinos at
redshift of $z<1000$, with an explicit justification on the use of a fluid approximation. 
We solve the collisionless Boltzmann equation in an
Einstein de-Sitter (EdS) universe, truncating the Boltzmann hierarchy at
$l_{\rm max}=1$ and $2$, and compare the resulting density contrast of neutrinos,
$\delta_{\rm\nu}^{\rm fluid}$, with that of the exact solutions of the Boltzmann equation that we derive in this paper.
Roughly speaking, the fluid approximation is accurate if neutrinos were
already non-relativistic when the neutrino density fluctuation of a given
wavenumber entered the horizon.
We find that the fluid approximation is accurate at few to 25\%
for massive neutrinos with $0.05<m_{\rm \nu}< 0.5~{\rm eV}$
at the scale of $k\lesssim0.4~h~{\rm Mpc^{-1}}$ and redshift of $z<10$.
This result quantifies the limitation of the fluid approximation,
for the massive neutrinos with $m_{\rm \nu}\lesssim 0.5~{\rm eV}$.
We also find that the density contrast calculated from fluid equations
(i.e., continuity and Euler equations) becomes a better approximation at a lower redshift, and the accuracy can be further improved by including an anisotropic stress term in the Euler equation. The anisotropic stress term effectively increases 
the pressure term by a factor of $9/5$.
\end{abstract}
\maketitle
%%%%%%%%%%%%%%%%%%%%%%%%%%%%%%%%%%%%%%%%%%%%%%%%%%%%%%%%%%%%%%%%%%
\section{Introduction}
What is the mass of neutrinos?
We know that at least two of three standard model neutrino species have
finite masses. The constraints on the squared mass differences of the
three species of neutrinos obtained from solar
\cite{davis:1994,itoh/etal:1996,cleveland/etal:1998,abdurashitov/etal:1999,hampel/etal:1999,altmann/etal:2000,abdurashitov/etal:2002,ahmad/etal:2002a,ahmad/etal:2002b,fukuda/etal:2002,ahmed/etal:2004,altmann/etal:2005,hosaka/etal:2006}
and atmospheric oscillation experiments
\cite{fukuda/etal:1998,surdo:2002,sanchez/etal:2003,ashie/etal:2005,hosaka/etal:2006b}
(reviews can be found in \cite{maltoni/etal:2004,lesgourgues/pastor:2006,gonzalez/maltoni:2008,schwetz/tortola/valle:2008}) are ($3~\sigma$ errors)
\begin{eqnarray}
\Delta m_{21}^2=\left(7.65^{+0.69}_{-0.60}\right)\times 10^{-5}~{\rm eV^2},\\
\left|\Delta m_{31}^2\right|=\left(2.40^{+0.35}_{-0.33}\right)\times 10^{-3}~{\rm eV^2}.
\end{eqnarray}
Therefore, the lower limit on the sum of neutrino masses is $0.058~{\rm eV}$.
Observations of the CMB and large-scale structure of the universe can provide
limits on the {\it absolute} mass of neutrinos. The current upper bounds on the
sum of neutrino masses are $\simeq 0.3-0.6~{\rm eV}$
\cite{mantz/allen/rapetti:2009,thomas/abdalla/lahav:2009,vikhlinin/etal:2009,hannestad/etal:2010,reid/etal:2010,sekiguchi/etal:2010}.
In this paper, we use $0.58~{\rm eV}$ ($95\%$ CL) from WMAP7yr
as a conservative upper bound on the sum of neutrino masses \cite{komatsu/etal:2010}.

The large-scale structure of the universe is a sensitive probe of neutrino masses
\cite{bond/efstathiou/silk:1980,doroshkevich/etal:1980,doroshkevich/etal:1980b,doroshkevich/khlopov:1981,doroshkevich/etal:1981,hu:1998,hu/eisenstein:1998,valdarnini/kahniashvili/novosyadlyj:1998,eisenstein/hu:1999,lewis/challinor:2002,lesgourgues/pastor:2006}.
Massive neutrinos suppress the small-scale matter power spectrum
by their large velocity dispersion,
The fractional amount of the suppression in the small-scale limit is roughly
given by
\begin{eqnarray}
\left|{\frac{\Delta P(k)}{P(k)}}\right|\simeq 8\frac{\Omega_{\rm\nu}}{\Omega_m},
\end{eqnarray}
with
\begin{eqnarray}
\Omega_{\rm\nu}h^2=\frac{\sum m_{\rm \nu,i}}{94.1~{\rm eV}},
\end{eqnarray}
where the summation is taken over the $i$-th species of neutrinos
\cite{hu/eisenstein:1998,hu/eisenstein/tegmark:1998,takada/komatsu/futamase:2006}.

Relativistic neutrinos are not a fluid.
Massive neutrinos, being collisionless, are not a fluid, either.
However, when the velocity dispersion of massive neutrinos becomes low enough,
they may be approximately treated as a fluid, just as we normally treat
Cold Dark Matter (CDM) particles as a fluid on large-scales.
While this is a reasonable expectation, as far as we know, the extent
to which the fluid approximation is valid for massive neutrinos
has not been discussed in the literature.

Then, why is a fluid approximation useful while we have Boltzmann
codes such as CMBfast \cite{seljak/zaldarriaga:1996} and CAMB \cite{lewis/challinor/lasenby:2000}, which solve the Boltzmann equations
numerically to the accuracy of order $\sim 0.1\%$?
First, we will have more physical insight to the growth of the neutrino density
fluctuations by directly solving continuity and Euler equations rather
than numerically solving a set of infinite order of Boltzmann hierarchy.
Second, and most importantly, as the density fluctuations become non-linear,
i.e., $\delta\sim 1$, we need to use higher-order perturbation theories
to accurately model the small-scale density fluctuations.
Since the higher-order perturbation theories have been constructed for CDM
with a fluid approximation, we cannot simply modify theories to include
massive neutrinos if a fluid approximation is not valid for those
particles. On the other hand, if a fluid approximation is valid for
some range of redshifts, length scales and neutrino masses, we can greatly
simplify the model of non-linear density fluctuations in the presence of
massive neutrino, as shown in \cite{shoji/komatsu:2009}.

In this paper, we shall study the validity of a fluid approximation
of massive neutrinos.
To achieve this goal, we first solve the Boltzmann equations describing the
evolution of the perturbed phase-space distribution function of massive
neutrinos {\it exactly} and compare the exact results to the results
with the fluid approximation, i.e., solutions with the higher multipole
moments ($l\ge 3$) ignored.
Then, we shall examine the ranges of applicability of fluid approximation
in both spatial and time scales, as a function of neutrino masses.

The rest of this paper is organized as follows.
In \S~\ref{sec:freestreaming}, we briefly review the effects of massive neutrino
free-streaming on the structure formation of the universe.
In \S~\ref{sec:boltzmann_hierarchy}, we provide the basic fluid equations
and the linearized Boltzmann equation required for our theoretical flame work.
In \S~\ref{sec:analytic}, we briefly discuss the analytic solutions of the
Boltzmann equation for collision-less particles.
In \S~\ref{sec:fluid_approx}, we compare the exact solutions of the
Boltzmann equations with the fluid approximation, and discuss the limitation
of the fluid approximation for several masses of massive neutrino.
Finally, in \S~\ref{sec:conclusions}, we discuss the implications
of our results and conclude.
In Appendix \ref{sec:sigma_app}, we discuss how to define the free-streaming
scale starting from the fluid equations.
In Appendix \ref{sec:exact_solution_app}, we give the detailed derivation of the exact solution
of the Boltzmann equation both for massless and massive neutrinos.
Even though our main interest is in massive neutrinos, our results
shown here are also applicable to collision-less particles in general, whose time
evolution of the perturbed phase space distribution follows the linearized
collision-less Boltzmann equation with the zero-th order phase space
distribution function being frozen at sufficiently early time
(i.e., we set the initial conditions of the neutrino transfer function
after the decoupling of neutrino, $\sim 1 ~{\rm MeV}$).
%%%%%%%%%%%%%%%%%%%%%%%%%%%%%%%%%%%%%%%%%%%%%%%%%%%%%%%%%%%%%%%%%%
\section{The Free-Streaming of the Massive Neutrino}
\label{sec:freestreaming}
We are interested in the mass range of $0.05<m_{\rm \nu,i}<0.58~{\rm eV}$
for the most massive species of neutrinos, which became non-relativistic well
after the matter radiation equality.
The mass density of the massive neutrinos relative to the total matter density
is given by
\begin{eqnarray}
f_{\rm\nu}\equiv\frac{\Omega_{\rm\nu}h^2}{\Omega_mh^2}=\frac{1}{\Omega_mh^2}
\frac{\sum_{\rm i}m_{\rm\nu,i}}{94.1{\rm eV}},
\label{eq:fn}
\end{eqnarray}
where the summation is taken over the different species of neutrinos.
Neutrinos become non-relativistic when the mean energy per particle
of neutrinos in the relativistic limit,
\begin{eqnarray}
\left<E\right>&\equiv&\frac{\int d^3p~p~
(\exp[p/T_{\rm\nu}(z)]+1)^{-1}}{\int d^3p~(\exp[p/T_{\rm\nu}(z)]+1)^{-1}}
\nonumber\\
&=&\frac{7\pi^4}{180\zeta(3)}T_{\rm\nu}\simeq 3.15T_{\rm\nu},
\end{eqnarray}
falls below $m_{\rm\nu,i}$.
By solving $3.15T_{\rm\nu,0}(1+z_{\rm nr})=m_{\rm\nu,i}$, one finds the redshift of
relativistic to non-relativistic transition epoch, $z_{\rm nr}$, as
\begin{eqnarray}
1+z_{\rm nr,i}\simeq1890\left(\frac{m_{\rm\nu,i}}{1{\rm eV}}\right),
\label{eq:znr}
\end{eqnarray}
for the $i$-th neutrino species.

The density fluctuation of neutrinos cannot grow within the horizon size until
neutrinos become non-relativistic. Once neutrinos become non-relativistic,
the neutrino density fluctuation begins to grow on scale greater than the so called
``free-streaming scale,'' which is set by the velocity dispersion of neutrinos:
\begin{eqnarray}
\sigma^2_{\rm\nu,i}(z)&\equiv&\frac{\int d^3p\ p^2/m_{\rm\nu,i}^2
(\exp[p/T_{\rm\nu}(z)]+1)^{-1}}{\int d^3p~(\exp[p/T_{\rm\nu}(z)]+1)^{-1}}
\nonumber\\
&=&\frac{15\zeta(5)}{\zeta(3)}\left(\frac4{11}\right)^{\frac23}
\frac{T^2_{\rm\gamma,0}(1+z)^2}{m^2_{\rm\nu,i}},
\label{eq:sigsq}
\end{eqnarray}
where $p$ is the proper momentum of the massive neutrino (see Appendix of
\cite{takada/komatsu/futamase:2006}).

The wavenumber corresponding to the free-streaming scale, $k_{\rm FS}$,
is defined by the single-fluid continuity and Euler equations:
\begin{eqnarray}
&&\dot{\delta}(\mathbf{k},\tau)+\theta(\mathbf{k},\tau)=0\\
&&\dot{\theta}(\mathbf{k},\tau)+\mathcal{H}(\tau)\theta(\mathbf{k},\tau)
+\left[\frac32\mathcal{H}^2(\tau)-k^2c_{\rm s}^2(\tau)\right]\delta(\mathbf{k},\tau)=0,
\nonumber\\
\end{eqnarray}
where
\footnote{
Here, we say $c_{\rm s}\simeq\sigma_{\rm\nu,i}$; however,
strictly speaking, the velocity dispersion defined in Eq.(\ref{eq:sigsq})
should not be used to define the free-streaming scale, $k_{\rm FS}$, as the
Euler equation contains sound speed, $c_{\rm s}^2\equiv\frac{\delta P}{\delta\rho}$,
not the velocity dispersion. In the non-relativistic limit, we have
$c_s=\frac{\sqrt{5}}{3}\sigma_{\rm\nu,i}\simeq 0.745\sigma_{\rm\nu,i}$.
We derive this relation in Appendix \ref{sec:sigma_app}.
}
\begin{eqnarray}
&&k_{\rm FS,i}(z)\equiv\sqrt{\frac32}\frac{\mathcal{H}(z)}{c_{\rm s}(z)}
\simeq\sqrt{\frac32}\frac{\mathcal{H}(z)}{\sigma_{\rm\nu,i}(z)}
\nonumber\\
&&\simeq\frac{0.677}{(1+z)^2}\left(\frac{m_{\rm\nu,i}}{1~{\rm eV}}\right)
[\Omega_m(1+z)^3+\Omega_{\Lambda}]^{\frac12}~h~{\rm Mpc^{-1}}.
\nonumber\\
\label{eq:kfs}
\end{eqnarray}
Here, derivatives are with respect to a conformal time, $d\tau=dt/a$,
$\mathcal{H}(\tau)\equiv \frac{\dot{a}(\tau)}{a(\tau)}$, and
$\theta(\mathbf{k},\tau)$ is a velocity divergence of the fluid.
Note that Eq.(\ref{eq:sigsq}) assumes that neutrinos are
non-relativistic.

%%%%%%%%%%%%%%%%%%%%%%%%%%%%%%%%%%%%%%%%%%%%%%%%%%%%%%%%%%%%%%%%%%
%insert figure here
\begin{figure*}[t]
\rotatebox{0}{%
  \includegraphics[width=8cm]{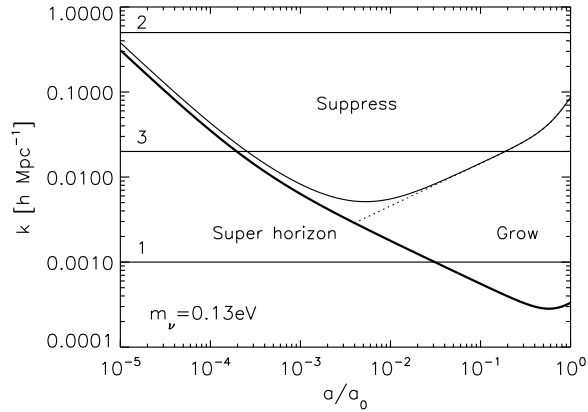}
}%
\caption{%
Free-streaming scale of a
massive neutrino, $k_{\rm FS,i}$, (black line), comoving horizon scale,
$aH(a)$, (thick black line) and an
approximation to the free-streaming scale in the non-relativistic limit
given by Eq.(\ref{eq:kfs}),
(dotted line) as functions of the scale factor, $a$.
We use $m_{\rm\nu,i}=0.13~{\rm eV}$.
The horizontal lines show (1) large, (2) small, and (3) intermediate
scale modes as described in \S~\ref{sec:freestreaming}.
}%
\label{kfs}
\end{figure*}
%%%%%%%%%%%%%%%%%%%%%%%%%%%%%%%%%%%%%%%%%%%%%%%%%%%%%%%%%%%%%%%%%%
In Figure \ref{kfs}, we show $k_{\rm FS,i}$ from Eq.(\ref{eq:kfs}) (dotted line),
comoving horizon scale, $aH(a)$, (thick solid line)
and $k_{\rm FS,i}$ calculated numerically from Eq.(\ref{eq:sigsq}),
where $m_{\rm\nu,i}$ is replaced by $\sqrt{p^2+m^2_{\rm\nu,i}}$ (thin solid line).
In this figure, we use $m_{\rm\nu,i}=0.13~{\rm eV}$.

We find that the free-streaming scale is close to the horizon size until the
relativistic to non-relativistic transition of a neutrino, and once the
neutrino becomes non-relativistic, the free-streaming scale decreases as
$k_{\rm FS}(a)\propto a^{1/2}$.
Let us examine the evolution of the neutrino density fluctuations at three
length scales:
\begin{enumerate}
\item At the large-scale, where $k\ll k_{\rm FS}(a)$ for all $a\le a_0$
($a_0$ is the present-day scale factor),
the neutrino density fluctuation starts to grow soon after the mode enters
the horizon, and its time evolution is identical to that of CDM,
$\delta_{\rm\nu}(k,a)=\delta_{\rm c}(k,a)$.
\item At the small-scale, where $k\gg k_{\rm FS}(a)$ for all $a\le a_0$,
the neutrino density fluctuation oscillates around
its initial value due to the free-streaming effect,
$\delta_{\rm\nu}(k,a)\sim\delta_{\rm\nu}(k,a_{\rm i})\simeq 0$.
\item At the intermediate scale, the mode first experiences the free-streaming
phase, and thus does not grow.
Once $k<k_{\rm FS}(a)$ is satisfied, the mode starts to grow, rapidly catching up
with the gravitational potential set up by CDM.
\end{enumerate}

\section{The Boltzmann Hierarchy and Fluid Approximation}
\label{sec:boltzmann_hierarchy}
In this section, we provide all the relevant equations and definitions
needed for our theoretical flame work, following \cite{ma/bertschinger:1995}
in the conformal-Newtonian gauge.

For fermions and bosons, we have the phase space distribution (in natural units)
given by
\begin{eqnarray}
f_0(q,\tau)&=&\frac{g_s}{(2\pi)^3}\frac1{e^{\epsilon(q,\tau)/aT(a)}\pm1},
\end{eqnarray}
where the sign of ``$+$'' is for fermions and ``$-$'' is for bosons,
$q$ and $\epsilon(q,\tau)\equiv\sqrt{q^2+a^2(\tau)m^2}$ are
the co-moving momentum (i.e., $q=a(\tau)p$) and the comoving energy
of a particle, respectively.
Here, $\tau$ is a conformal time, which is related to the proper time by
$d\tau=dt/a(t)$, and $g_s$ is a number of degrees of freedom.
The linear order perturbation to the distribution function,
$\Psi(\mathbf{k},\mathbf{\hat{n}},q,\tau)$, is defined as
\begin{eqnarray}
f(\mathbf{k},\hat{\mathbf{n}},q,\tau)&=&f_0(q,\tau)
[1+\Psi(\mathbf{k},\hat{\mathbf{n}},q,\tau)],
\end{eqnarray}
where $q\equiv|\mathbf{q}|$ and $\hat{\mathbf{n}}\equiv\mathbf{q}/q$.

Since neutrinos decoupled while they were highly relativistic,
the unperturbed distribution function after the neutrino decoupling
continues to be given by its relativistic form:
\begin{eqnarray}
f_0(q)&=&\frac{g_s}{(2\pi)^3}\frac1{e^{q/aT(a)}\pm1},
\end{eqnarray}
even after neutrinos become non-relativistic.
The temperature of such collision-less particles decreases as
$T(a)=T_0({a_0}/a)$,
even when they are non-relativistic.

The evolution of the linearized phase-space distribution for
collision-less particles such as CDM and neutrinos is
governed by the linearized collision-less Boltzmann equation,
\begin{eqnarray}
\frac{\partial\Psi(\mathbf{k},\mathbf{\hat{n}},q,\tau)}
{\partial\tau}+i\frac{q}{\epsilon(q,\tau)}
(\mathbf{k}\cdot\mathbf{\hat{n}})\Psi(\mathbf{k},\mathbf{\hat{n}},q,\tau)
\nonumber\\
+\frac{d\ln f_0(q)}{d\ln q}\left[
\dot{\phi}(k,\tau)-i\frac{\epsilon(q,\tau)}{q}(\mathbf{k}\cdot\mathbf{\hat{n}})\psi(k,\tau)
\right]=0,
\label{eq:boltzmann}
\end{eqnarray}
where $\psi$ and $\phi$ are a Newtonian gravitational potential and a curvature
perturbation, respectively.
\footnote{
In the original work of \cite{ma/bertschinger:1995}, $\psi$ and $\phi$ are
defined as scalar perturbations in the metric in the conformal Newtonian gauge:
$ds^2=a^2(\tau)[-(1+2\psi)d\tau^2+(1-2\phi)dx^idx_i]$.
They are related to the gauge invariant variables $\Phi_A$ and $\Phi_H$ of
\cite{bardeen:1980} and $\Psi$ and $\Phi$ of \cite{kodama/sasaki:1984} by
$\psi=\Phi_A=\Psi$ and $\phi=-\Phi_H=-\Phi$.
}

To simplify the equation, we define $\tilde{\Psi}(\mathbf{k},\hat{\mathbf{n}},q,\tau)
\equiv\Psi(\mathbf{k},\hat{\mathbf{n}},q,\tau)
\left(\frac{d\ln f_0(q)}{d\ln q}\right)^{-1}$,
and replace the time derivative from $\tau$ to $x\equiv k\tau$,
and re-write Eq.(\ref{eq:boltzmann}) as
\begin{eqnarray}
\frac{\partial\tilde{\Psi}(\mathbf{k},\mathbf{\hat{n}},q,x)}
{\partial x}+i\frac{q}{\epsilon(q,x)}\mu
\tilde{\Psi}(\mathbf{k},\mathbf{\hat{n}},q,x)+\frac{\partial\phi(k,x)}{\partial x}
\nonumber\\
-i\frac{\epsilon(q,x)}{q}\mu\psi(k,x)=0,
\label{eq:boltzmann2}
\end{eqnarray}
where $\mu$ is a cosine between the wavenumber and momentum, i.e.,
$\mathbf{k}\cdot\hat{\mathbf{n}}\equiv k\mu$.
Finally, we expand the Boltzmann equation (Eq.(\ref{eq:boltzmann2}))
by Legendre polynomials, using
\begin{eqnarray}
\tilde{\Psi}(\mathbf{k},\hat{\mathbf{n}},q,x)&=&\sum^{\infty}_{l=0}(-i)^l
(2l+1)\tilde{\Psi}_l(k,q,x)P_l(\mu),
\end{eqnarray}
and obtain a set of infinite series of differential equations (also known as
Boltzmann hierarchy) as follows:
\begin{eqnarray}
\tilde{\Psi}'_0(k,q,x)&=&-\frac{q}{\epsilon(q,x)}\tilde{\Psi}_1(k,q,x)-\phi'(k,x),
\label{eq:psi0}\\
\tilde{\Psi}'_1(k,q,x)&=&\frac{q}{3\epsilon(q,x)}
[\tilde{\Psi}_0(k,q,x)-2\tilde{\Psi}_2(k,q,x)]
\nonumber\\
&-&\frac{\epsilon(q,x)}{3q}\psi(k,x),
\label{eq:psi1}\\
\tilde{\Psi}'_l(k,q,x)&=&\frac{q}{(2l+1)\epsilon(q,x)}
[l\tilde{\Psi}_{l-1}(k,q,x)
\nonumber\\
&-&(l+1)\tilde{\Psi}_{l+1}(k,q,x)]
\ \ ({\rm for} ~l\ge2),\label{eq:psil}
\end{eqnarray}
where the primes denote derivatives with respect to $x\equiv k\tau$.
Here, $\tilde{\Psi}_0(k,q,x)$ is sourced by $\tilde{\Psi}_1(k,q,x)$. All the
successive multipoles with $l\ge1$, $\tilde{\Psi}_{l\ge1}(k,q,x)$, are sourced by
$\tilde{\Psi}_{l-1}(k,q,x)$ and $\tilde{\Psi}_{l+1}(k,q,x)$,
so that the evolution of $l$-th multipole propagates
the whole system of equations back and forth. In order to close the system of
equations, we need to truncate the Boltzmann hierarchy at some
finite multipole, $l_{\rm max}$.
Now, the question is, ``{\it in which condition the fluid approximation
(i.e., $l_{\rm max}=1$ or $2$) is valid?}''

To make a contact with the familiar form of fluid equations,
we relate multipoles of the perturbed distribution function,
$\Psi_l(k,q,\tau)$, to the quantities such as the
density contrast,
$\delta(k,\tau)\equiv\frac{\delta\rho(k,\tau)}{\bar{\rho}(\tau)}$,
velocity dispersion, $\theta(k,\tau)$, and anisotropic stress,
$\sigma(k,\tau)$, by integrating $\Psi_l(k,q,\tau)$
over the momentum space with appropriate powers of $q$:
\begin{eqnarray}
\bar{\rho}(\tau)\!\!&=&\!\!\frac{4\pi}{a^4(\tau)}\!\!\int\!\! q^2dq\ \epsilon(q,\tau)f_0(q),
\label{eq:rho_bar}\\
\bar{P}(\tau)\!\!&=&\!\!\frac{4\pi}{3a^4(\tau)}\!\!\int\!\! q^2dq\frac{q^2}{\epsilon(q,\tau)}
f_0(q),
\label{eq:P_bar}\\
\delta\rho(k,\!\tau\!)\!\!&=&\!\!\frac{4\pi}{a^4(\!\tau\!)}\!\!\int\!\! q^2dq\ \epsilon(q,\!\tau)f_0(q)
\Psi_0(k,\!q,\!\tau),
\label{eq:delta_rho}\\
\delta P(k,\!\tau\!)\!\!&=&\!\!\frac{4\pi}{3a^4(\!\tau\!)}\!\!\int\!\! q^2dq\frac{q^2}{\epsilon(q,\!\tau)}
f_0(q)\Psi_0(k,\!q,\!\tau),
\label{eq:delta_P}\\
(\bar{\rho}\!+\!\bar{P})\theta(k,\!\tau\!)\!\!&=&\!\!\frac{4\pi k}{a^4(\!\tau\!)}\!\!\int\!\! q^2dq\ q f_0(q)
\Psi_1(k,\!q,\!\tau),
\label{eq:theta}\\
(\bar{\rho}\!+\!\bar{P})\sigma(k,\!\tau\!)\!\!&=&\!\!\frac{8\pi}{3a^4(\!\tau\!)}\!\!\int\!\! q^2dq
\frac{q^2}{\epsilon(q,\!\tau)}f_0(q)\Psi_2(k,\!q,\!\tau).
\label{eq:sigma}
\end{eqnarray}
We obtain the fluid equations by truncating the Boltzmann hierarchy
at $l_{\rm max}=2$:
\begin{eqnarray}
\dot{\delta}(k,\tau)\!\!&=&\!\!-[1+w(\tau)][\theta(k,\tau)-3\dot{\phi}(k,\tau)]
\nonumber\\
&-&3\frac{\dot{a}(\tau)}{a(\tau)}\left[\frac{\delta P(k,\tau)}
{\delta\rho(k,\tau)}-w(\tau)\right]\delta(k,\tau),
\label{eq:continuity}\\
\dot{\theta}(k,\tau)\!\!&=&\!\!-\frac{\dot{a}(\tau)}{a(\tau)}[1-3w(\tau)]
\theta(k,\tau)-\frac{\dot{w}(\tau)}{1+w(\tau)}\theta(k,\tau)
\nonumber\\
&+&\frac{\delta P(k,\tau)/\delta\rho(k,\tau)}{1+w(\tau)}k^2\delta(k,\tau)
-k^2\sigma(k,\tau)+k^2\psi,\nonumber\\
\label{eq:Euler}\\
\dot{\sigma}(k,\tau)\!\!&=&\!\!-\frac{\dot{a}(\tau)}{a(\tau)}[2-3w(\tau)]
\sigma(k,\tau)-\frac{\dot{w}(\tau)}{1+w(\tau)}\sigma(k,\tau)
\nonumber\\
&+&\frac4{15}\Theta(k,\tau)+\frac{\dot{a}(\tau)}{a(\tau)}\Sigma(k,\tau),
\label{eq:anisotropic_strees}
\end{eqnarray}
where $w(\tau)\equiv \bar{P}(\tau)/\bar{\rho}(\tau)$ is an equation
of state, and we have defined the following variables:
\begin{eqnarray}
(\bar{\rho}+\bar{P})\Theta(k,\tau)\!\!&=&\!\!\frac{4\pi k}{a^4(\tau)}\int q^2dq\ q
\left(\frac{q}{\epsilon(q,\tau)}\right)^2
\nonumber\\
&\times&f_0(q)
\Psi_1(k,q,\tau),
\label{eq:theta_v2}\\
(\bar{\rho}+\bar{P})\Sigma(k,\tau)\!\!&=&\!\!\frac{8\pi}{3a^4(\tau)}\int q^2dq
\frac{q^2}{\epsilon(q,\tau)}\left(\frac{q}{\epsilon(q,\tau)}\right)^2
\nonumber\\
&\times&f_0(q)\Psi_2(k,q,\tau).
\label{eq:sigma_v2}
\end{eqnarray}
In the relativistic and non-relativistic limits, where majority of
neutrinos in the phase space distribution have momenta of
$q\sim\epsilon(q,\tau)$ and $q\ll\epsilon(q,\tau)$, we have
$(w,\dot{w},\frac{\delta P}{\delta\rho},\Theta,\Sigma)=(\frac13,0,\frac13,\theta,\sigma)$ and $(0,0,0,0,0)$, respectively.

Since CDM is non-relativistic throughout the redshift
of our interest, we can greatly simplify the calculation of the density contrast
of CDM by fluid approximation (i.e., $l_{\rm max}=1$).
We have
\begin{eqnarray}
\dot{\delta}(k,\tau)\!\!&=&\!\!-\theta(k,\tau)+3\dot{\phi}(k,\tau),
\\
\dot{\theta}(k,\tau)\!\!&=&\!\!-\frac{\dot{a}(\tau)}{a(\tau)}
\theta(k,\tau)+k^2\psi(k,\tau).
\end{eqnarray}
As for massive neutrinos, the fluid approximation may or may not be valid,
depending on the mass, scale or redshift of interest.
We check the validity of the fluid approximation for massive neutrinos
by comparing to the exact solutions in section \S~\ref{sec:fluid_approx}.

When $l_{\rm max}=2$, Eq.(\ref{eq:psi0}) and (\ref{eq:psil}) give a useful relation
between $\tilde{\Psi}_0(k,q,x)$ and $\tilde{\Psi}_2(k,q,x)$:
\begin{eqnarray}
\frac{\partial}{\partial x}\left[\tilde{\Psi}_2(k,q,x)+
\frac25\tilde{\Psi}_0(k,q,x)+\frac25\phi(k,x)\right]=0,
\label{eq:ansatz}
\end{eqnarray}
which gives
\begin{eqnarray}
&&\tilde{\Psi}_2(k,q,x)+\frac25\tilde{\Psi}_0(k,q,x)+\frac25\phi(k,x)
\nonumber\\
&&=\tilde{\Psi}_2(k,q,x_i)+\frac25\tilde{\Psi}_0(k,q,x_i)+\frac25\phi(k,x_i),
\end{eqnarray}
where $x_i\ll 1$ is an initial time.
With this relation and Eqs.(\ref{eq:delta_P}) and (\ref{eq:sigma}),
we can rewrite the anisotropic stress, $\sigma(k,\tau)$,
in the Euler equation (Eq.(\ref{eq:Euler})) in terms of pressure, $\delta P(k,\tau)$,
as
\begin{eqnarray}
&&k^2\sigma(k,\tau)+\frac45\frac{\delta P(k,\tau)/\delta\rho(k,\tau)}{1+w(\tau)}k^2\delta(k,\tau)=const,\nonumber\\
\label{eq:sigtopressure0}
\end{eqnarray}
where we have set $\phi=const$.
At late times, $\tau\gg\tau_i$, where
$\delta(k,\tau)\gg\delta(k,\tau_i)$
and $\sigma(k,\tau)\gg\sigma(k,\tau_i)$, the right hand side of
Eq.(\ref{eq:sigtopressure0}) is negligible compared to the second term
on the left hand side.
Therefore, we have
\begin{eqnarray}
&&k^2\sigma(k,\tau)\simeq-\frac45\frac{\delta P(k,\tau)/\delta\rho(k,\tau)}{1+w(\tau)}k^2\delta(k,\tau),\nonumber\\
\label{eq:sigtopressure}
\end{eqnarray}
This result shows that $\sigma$ increases the pressure by a factor of $\frac95$.
The free-streaming scale, $k_{\rm FS}$, is then reduced by a factor of
$\frac{3}{\sqrt{5}}\sim 1.34$.
%%%%%%%%%%%%%%%%%%%%%%%%%%%%%%%%%%%%%%%%%%%%%%%%%%%%%%%%%%%%%%%%%%
\section{Analytic Solutions for the Boltzmann Equation}
\label{sec:analytic}
In this section, we briefly describe analytic solutions of the
Boltzmann equation (Eq.(\ref{eq:boltzmann2})), to which the fluid approximation is compared. We give a detailed derivation of the solutions in Appendix \ref{sec:exact_solution_app}.

Instead of expanding the Boltzmann equation
by Legendre polynomials as in \cite{ma/bertschinger:1995},
we first find a formal solution of Eq.(\ref{eq:boltzmann2}):
\begin{eqnarray}
\tilde{\Psi}(k,q,\mu,x)&=&\tilde{\Psi}(k,q,\mu,x_{i})e^{-i\mu[z(x)-z(x_i)]}
\nonumber\\
&+&\int^{x}_{x_i}dx'e^{-i\mu[z(x)-z(x')]}S(k,q,\mu,x'),
\label{eq:boltzmann_soln}
\end{eqnarray}
where
\begin{eqnarray}
S(k,q,\mu,x)\equiv i\frac{\epsilon(q,x)}{q}\mu\psi(k,x)-\frac{\partial\phi(k,x)}
{\partial x},
\label{eq:source_term}
\end{eqnarray}
and we have defined
\begin{eqnarray}
z(x)\equiv\int^x_C\frac{q}{\epsilon(q,x')}dx'.
\label{eq:z}
\end{eqnarray}
The lower integration boundary, $C\in\mathcal{R}$, is an arbitrary constant.
We then expand the above formal solution with Legendre
polynomials. We will need the following relation of the Wigner 3-$j$ symbols,
\begin{eqnarray}
\int^{1}_{-1}\frac{d\mu}{2}P_{l}(\mu)P_{l'}(\mu)P_{l''}(\mu)=
\begin{pmatrix}
l & l' & l''\\
0 & 0 & 0
\end{pmatrix}^2.
\end{eqnarray}
The solution for $\tilde{\Psi}_l(k,q,x)$ for a given $l$ is given by
\begin{eqnarray}
&&\tilde{\Psi}_{l}(k,q,x)=
\sum_{l'}\sum_{l''}(-i)^{l'+l''-l}(2l'+1)(2l''+1)
\nonumber\\
&&\times\tilde{\Psi}_{l'}(k,q,x_i)
j_{l''}(z-z_i)
\begin{pmatrix}
l & l' & l''\\
0 & 0 & 0
\end{pmatrix}^2
\nonumber\\
&-&\!\!\psi(k)\!\!\int^x_{x_i}\!\!\!\!dx'\frac{\epsilon(q,x')}{q}
\!\left[
\frac{l}{2l+1}j_{l-1}(z\!-\!z')
-\frac{l+1}{2l+1}j_{l+1}(z\!-\!z')
\right],
\nonumber\\
\label{eq:psi_l_massive}
\end{eqnarray}
where we have assumed $\dot{\psi}(k,x)=\dot{\phi}(k,x)=0$
(which is satisfied in an Einstein de-Sitter, EdS, universe).
We derive solutions for the most general case
(i.e., $\dot{\psi}(k,x)\ne 0$ and $\dot{\phi}(k,x)\ne 0$)
in Appendix \ref{sec:exact_solution_app}.

As we see in Eq.(\ref{eq:psi_l_massive}), the infinite series of the
Boltzmann hierarchy, Eqs.(\ref{eq:psi0})$\sim$(\ref{eq:psil}), is now expressed
in terms of the spherical Bessel functions, $j_l(x)$, its integrals weighted by
$\epsilon(q,x)/q$, and the infinite sum of the initial values of
$\tilde{\Psi}_l(k,q,x)$.

While Eq.(\ref{eq:psi_l_massive}) appears to have infinite sums over $l$,
the sum actually truncates because, at initial time (which is taken to be
before the horizon re-entry), $\tilde{\Psi}_l(k,q,x)$ for $l\ge3$ can be
ignored \cite{ma/bertschinger:1995}.
Together with the triangular inequalities of the Wigner 3-$j$ symbols,
\begin{eqnarray}
|l-l'|\le l''\le l+l',
\end{eqnarray}
only finite terms remain in the solution of $\tilde{\Psi}_l(k,q,x)$.

The explicit solutions of $\tilde{\Psi}_0(k,q,x)$ and
$\tilde{\Psi}_1(k,q,x)$ are
\begin{eqnarray}
&&\tilde{\Psi}_0(k,q,x)=\tilde{\Psi}_0(k,q,x_i)j_0(z-z_i)
\nonumber\\
&&-3\tilde{\Psi}_1(k,q,x_i)j_1(z-z_i)
+5\tilde{\Psi}_2(k,q,x_i)j_2(z-z_i)
\nonumber\\
&&+\psi(k)\int^x_{x_i}dx'\frac{\epsilon(q,x')}{q}j_1(z-z'),
\label{eq:psi0_analytic}\\
&&\tilde{\Psi}_1(k,q,x)=\tilde{\Psi}_0(k,q,x_i)j_1(z-z_i)
\nonumber\\
&&+\tilde{\Psi}_1(k,q,x_i)j_0(z-z_i)
-2\tilde{\Psi}_2(k,q,x_i)j_1(z-z_i)
\nonumber\\
&&-2\tilde{\Psi}_1(k,q,x_i)j_2(z-z_i)
+3\tilde{\Psi}_2(k,q,x_i)j_3(z-z_i)
\nonumber\\
&&-\psi(k)\int^x_{x_i}dx'\frac{\epsilon(q,x')}{q}\left[\frac13j_0(z-z')
-\frac23j_2(z-z')\right].
\nonumber\\
\label{eq:psi1_analytic}
\end{eqnarray}

Let us examine the  behavior of $\tilde{\Psi}_l(k,q,x)$.
At sufficiently late time, $z(x)\gg z(x_i)$, all the terms
containing initial values of $\tilde{\Psi}_l(k,q,x)$ become
negligible, as $j_l(z)\to 0$ for $z\gg l$.
The last term, which does not depend on initial values, is the dominant term.
For relativistic neutrinos, $\epsilon(q,x)=q$, the last term is proportional
to $\int^x_{x_i}dx'j_l(x-x')$, which approaches constant for $x\gg l$.
For non-relativistic neutrinos, $\epsilon(q,x)=a(\tau)m\propto x^2$,
$j_l(z)$ does not oscillate for $x\gg 1$ (because
$z(x)=\int\frac{q}{a(x)m}dx\propto\frac1{x}$).
Thus, the integrand, $\frac{\epsilon(q,x)}{q}j_l(z)$, grows.

%%%%%%%%%%%%%%%%%%%%%%%%%%%%%%%%%%%%%%%%%%%%%%%%%%%%%%%%%%%%%%%%%%
\section{The Validity of the Fluid Approximation}
\label{sec:fluid_approx}
Before we start, let us remember why fluid approximation may be valid
for massive particles. Eq.(\ref{eq:psi1}) shows that the $l=2$ mode
becomes unimportant when the gravitational force (the last term)
becomes dominant. The ratio of the last term to the first two terms
is of order $(\epsilon/q)^2$, which is unity for relativistic particles,
but it is much greater than unity for non-relativistic particles.
Thus, $l\ge 2$ modes become irrelevant for the evolution of $l=0$ and $1$
modes, allowing us to truncate the Boltzmann hierarchy at $l_{\rm max}=1$.

For example, when the distribution function of neutrinos is dominated by the
non-relativistic states (i.e., $q\ll a(x)m$), we have the following solutions for
$\tilde{\Psi}_0(k,q,x)$ and $\tilde{\Psi}_1(k,q,x)$ with constant $\phi$ and $\psi$,
\begin{eqnarray}
\tilde{\Psi}_0(k,q,x)\!\!&=&\!\!\frac{\psi}{18}\left[x\left(4\frac{k}{C}\!+\!x\right)
\!-\!16\left(\frac{k}{C}\right)^2\ln\left(4\frac{k}{C}\!+\!x\right)\right],
\nonumber\\
\label{eq:psi0_asymp}\\
\tilde{\Psi}_1(k,q,x)\!\!&=&\!\!-\frac{x^2}{36}\left(\frac{C}{k}\right)^2\frac{m}{q}
\left(6\frac{k}{C}+x\right)\psi,
\label{eq:psi1_asymp}
\end{eqnarray}
where $C$ is a constant, and the fastest growing modes grow as
$\tilde{\Psi}_0(k,q,x)\propto a(x)$
and $\tilde{\Psi}_1(k,q,x)\propto a(x)\sqrt{1+a(x)}$.

The observable quantities such as
$\delta(k,x)$ and $\theta(k,x)$ are given by the integrals of
$\tilde{\Psi}_0(k,q,x)$ and $\tilde{\Psi}_1(k,q,x)$. They pick up
contributions from the relativistic particles ($\epsilon(q,x)\sim q$)
as well, but the phase-space number density of those relativistic particles
is exponentially suppressed.
To see this, we rewrite $f_0$ of relativistic particles as
\begin{eqnarray}
f_0(q)=\frac{g_s}{(2\pi)^3}\frac1{e^{\frac{q}{m}\frac{m}{aT(a)}}\pm1},
\label{eq:f0_schematic}
\end{eqnarray}
where $\frac{q}{m}>1$ for relativistic particles in the phase space distribution.

For CDM with $m\sim 1~{\rm GeV}$, the mass to temperature ratio,
$m/aT(a)$, is very large for the time that is relevant to the structure formation;
thus, only extremely non-relativistic particles, $q/m\ll1$,
contribute to the density contrast, $\delta(k,\tau)$.
As neutrinos are much lighter than CDM, relativistic particles
may or may not contribute to the density contrast significantly.
For example, to calculate the massless neutrino density contrast accurately, 
we need to calculate the $\tilde{\Psi}_0(k,q,x)$ for relativistic particles
including all the higher multipoles in principle.
If the neutrinos are sufficiently massive, then the last term in Eq.(\ref{eq:psi1})
becomes dominant over the first two terms for most of the particles in the
phase-space distribution, $f_0(q)$, making the fluid approximation
(i.e., $l_{\rm max}=1$ or $2$) valid.

Note that a fluid approximation does not imply that all the higher moments of the
Boltzmann equations are small. It just means that the evolution equations of
$\delta(k,x)$ and $\theta(k,x)$ are decoupled from the higher moments
of the Boltzmann equations.

\subsection{$\tilde{\Psi}_l(k,q,x)$ with various $l_{\rm max}$}
\label{sec:psi_l}
To check the validity of fluid approximation, we need to compare the exact solutions
of $\tilde{\Psi}_0(k,q,x)$ and $\tilde{\Psi}_1(k,q,x)$ to the approximate
solutions of $\tilde{\Psi}_0(k,q,x)$ and $\tilde{\Psi}_1(k,q,x)$
with finite $l_{\rm max}=1$, $2$ and $3$.
To simplify the problem, we make the following (reasonable) approximations:
\begin{enumerate}
\item The universe is flat and matter dominated (EdS), for which
$\dot{\phi}=\dot{\psi}=0$.
\item We ignore the evolution of $\phi$ and $\psi$ caused by massive neutrinos,
which is a good approximation for proving the validity of the fluid approximation.
That is, $\psi$ is determined by CDM only.
\item The initial perturbations are adiabatic and the wave lengths of the initial
perturbations are greater than the horizon size, i.e., $x_i\ll 1$.
Specifically, we have (Eq.(97) of \cite{ma/bertschinger:1995}):
\end{enumerate}
\begin{eqnarray}
\tilde{\Psi}_0(k,q,x_i)&=&-\frac14\delta_{\rm\nu}(k,x_i)=\frac12\psi,
\label{eq:psi0_ini}\\
\tilde{\Psi}_1(k,q,x_i)&=&-\frac13\frac{\epsilon(q,x_i)}{q}
\frac{\theta_{\rm\nu}(k,x_i)}{k}
\nonumber\\
&=&-\frac16\frac{\epsilon(q,x_i)}{q}x_i\psi,
\label{eq:psi1_ini}\\
\tilde{\Psi}_2(k,q,x_i)&=&-\frac12\sigma_{\rm\nu}(k,x_i)
=-\frac1{30}x_i^2\psi,
\label{eq:psi2_ini}\\
\tilde{\Psi}_{l\ge 3}(k,q,x_i)&=&0,\label{eq:psil_ini}
\end{eqnarray}
where the most of neutrinos in the phase-space distribution are initially
highly relativistic, $\epsilon(q,x_i)\sim q$.

We rewrite the Boltzmann equations (Eqs.(\ref{eq:psi0}) $\sim$ (\ref{eq:psil}))
in terms of dimensionless parameters $k/C$ and $m/q$ defined below:
the ratio of the comoving energy to the comoving momentum of a particle is given as
\begin{eqnarray}
\frac{\epsilon(q,x)}{q}=\sqrt{1+a^2(x)\left(\frac{m}{q}\right)^2},
\end{eqnarray}
and the scale factor is given as
\begin{eqnarray}
a(x)=\frac{C}{k}x+\left(\frac{C}{k}\frac{x}{2}\right)^2,
\end{eqnarray}
where
\begin{eqnarray}
C=\sqrt{\frac{8\pi G\bar{\rho}_m(a_{\rm eq})}{3}}
&=&H_0\sqrt{\Omega_m}\left(\frac{a_{eq}}{a_0}\right)^{-\frac12}\nonumber\\
&=&0.0183\sqrt{\Omega_m}~h~{\rm Mpc^{-1}},
\end{eqnarray}
with $a_{\rm eq}\equiv 1$, $\bar{\rho}_{eq}=\bar{\rho}(a_{\rm eq})$,
and we assume the matter radiation equality to happen at
$1+z_{\rm eq}=a_0/a_{\rm eq}=3000$.
For $\Omega_r=8.47\times 10^{-5}$, $\Omega_m=0.25$ and $C=0.0092h~{\rm Mpc^{-1}}$.

With this convention for the scale factor, given comoving momentum is equal to
the physical momentum (i.e., $q=p$) at the matter radiation equality,
$a=a_{eq}$, and therefore, Eq.(\ref{eq:f0_schematic}) becomes,
\footnote{%
The original paper did not take this factor
of $a_{eq}/a_0$ into account,
and therefore, neutrino mass was overestimated by $a_0/a_{eq}=3000$.
This correction does not affect any qualitative/quantitative argument for
$\tilde{\Psi}_l$, but changes the interpretations of the resulting
$\delta_{\nu}$ and $\theta_{\nu}$.
}%

\begin{eqnarray}
f_0(q)&=&\frac{g_s}{(2\pi)^3}\frac1{e^{\frac{q}{m}\frac{m}{T_{eq}}}\pm1}\nonumber\\
&=&\frac{g_s}{(2\pi)^3}\frac1{e^{\frac{q}{m}\frac{m}{T_0}\frac{a_{eq}}{a_0}}\pm1}.
\label{eq:f0_mod}
\end{eqnarray}
%%%%%%%%%%%%%%%%%%%%%%%%%%%%%%%%%%%%%%%%%%%%%%%%%%%%%%%%%%%%%%%%%%
%insert figure here
\begin{figure*}[t]
\rotatebox{0}{%
  \includegraphics[width=14cm]{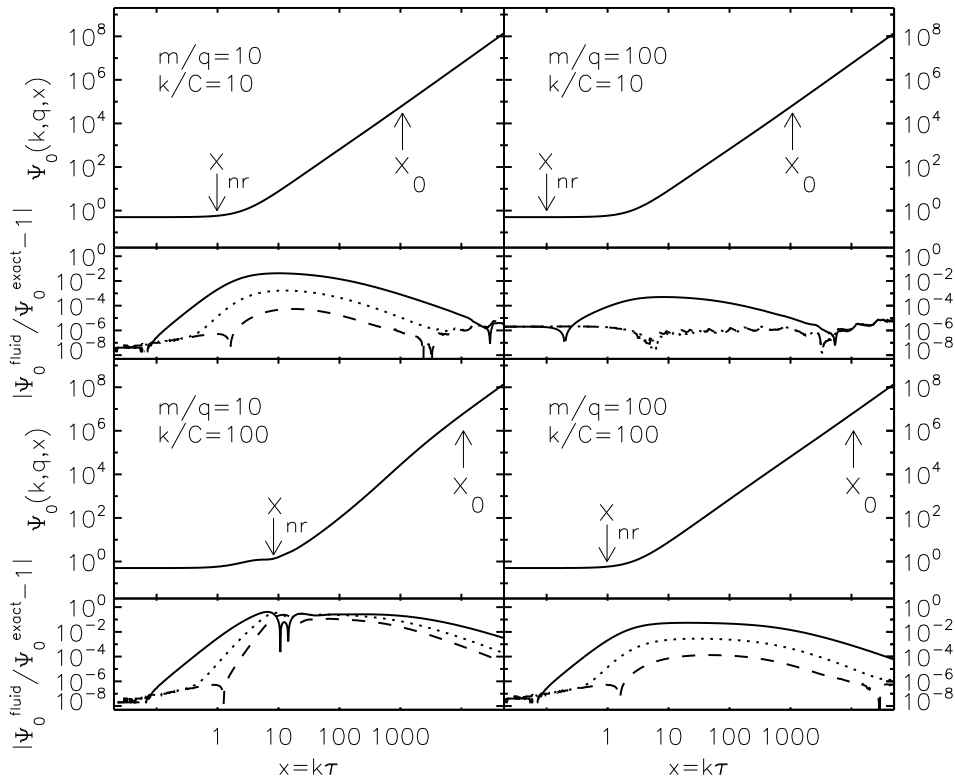}
}%
\caption{%
We show $\tilde{\Psi}_0(k,q,x)$ as functions of $x\equiv k\tau$ with two different
scales ($k/C=10$ and $100$, where $C=0.0092~h~{\rm Mpc^{-1}}$) and two different momenta
($m/q=10$ and $100$). $\tilde{\Psi}_0(k,q,x)$ is calculated from the exact solution, and
the fractional difference is given as
$\Delta\tilde{\Psi}_0/\tilde{\Psi}_0\equiv\tilde{\Psi}_0^{\rm fluid}/\tilde{\Psi}_0^{\rm exact}-1$,
where the solid line is for $l_{\rm max}=1$, the dotted line is for $l_{\rm max}=2$
and the dashed line is for $l_{\rm max}=3$.
}%
\label{psi0}
\end{figure*}
%%%%%%%%%%%%%%%%%%%%%%%%%%%%%%%%%%%%%%%%%%%%%%%%%%%%%%%%%%%%%%%%%%
%insert figure here
\begin{figure*}[t]
\rotatebox{0}{%
  \includegraphics[width=14cm]{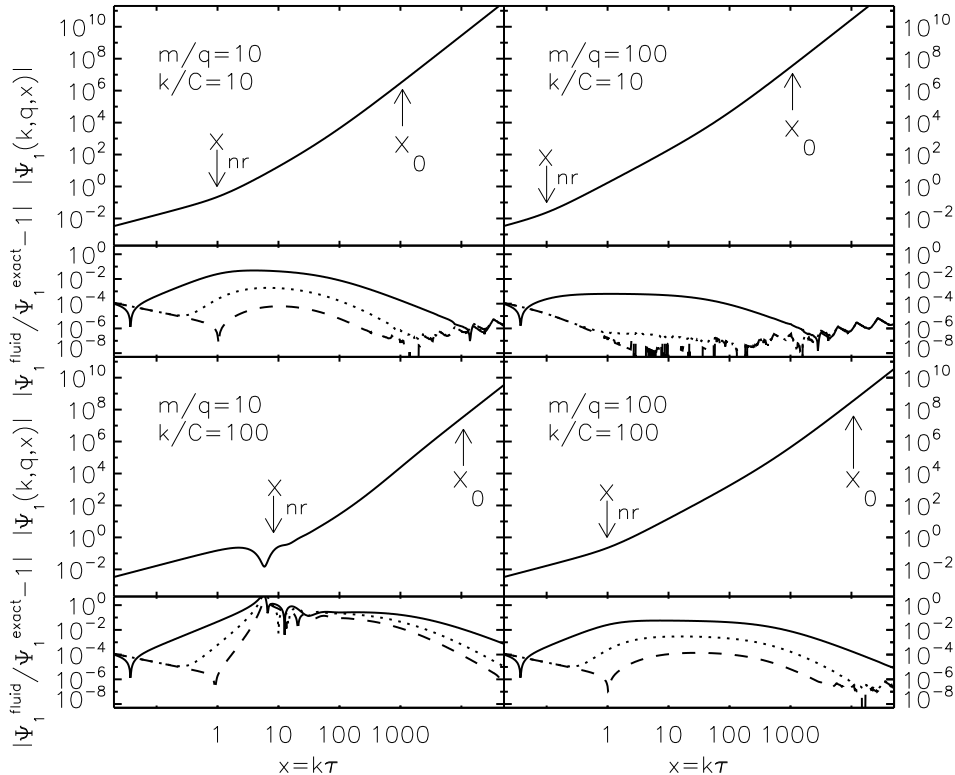}
}%
\caption{%
Same as Figure \ref{psi0} for $\tilde{\Psi}_1(k,q,x)$.
}%
\label{psi1}
\end{figure*}
%%%%%%%%%%%%%%%%%%%%%%%%%%%%%%%%%%%%%%%%%%%%%%%%%%%%%%%%%%%%%%%%%%
%insert figure here
\begin{figure*}[t]
\rotatebox{0}{%
  \includegraphics[width=14cm]{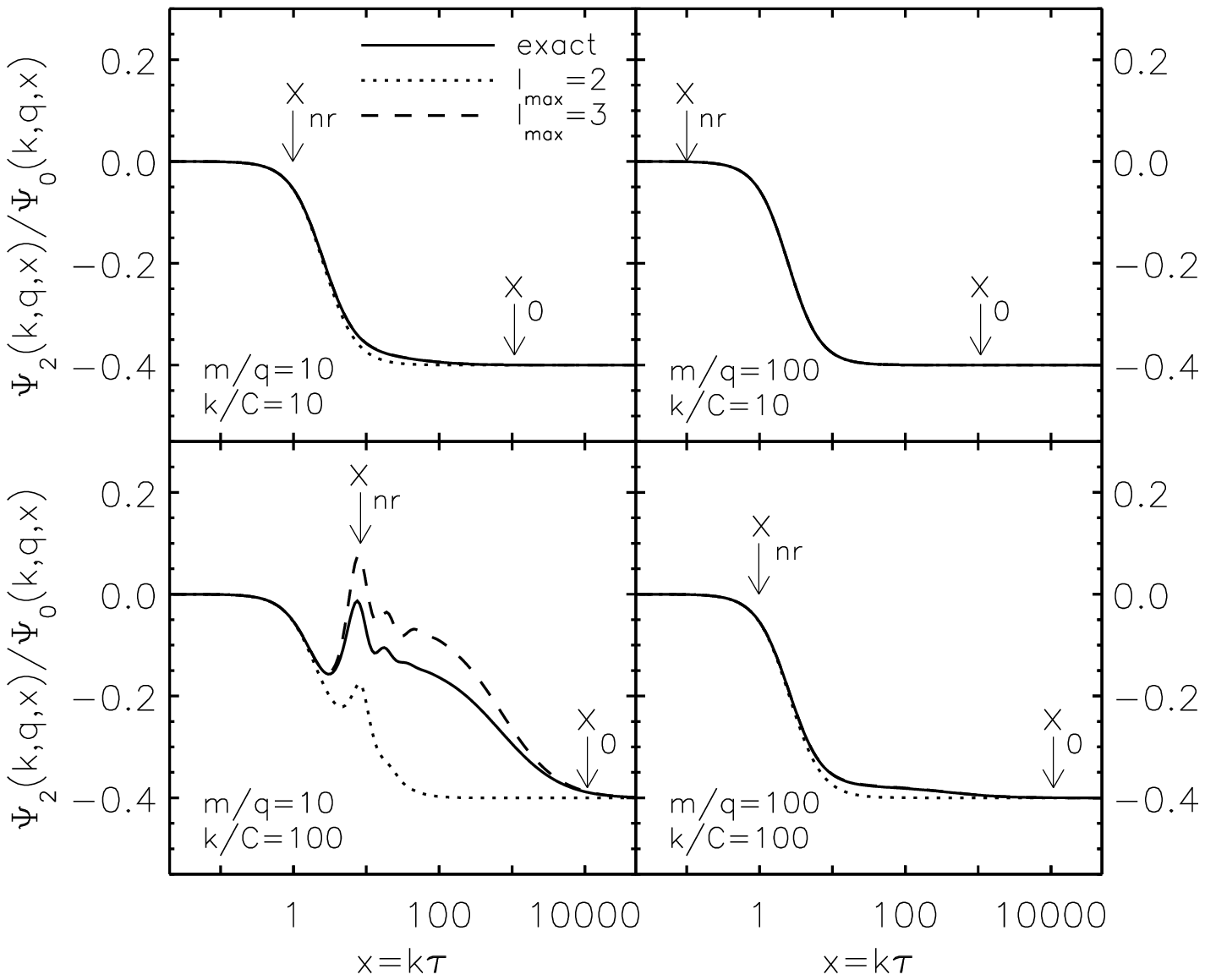}
}%
\caption{%
We show $\tilde{\Psi}_2(k,q,x)/\tilde{\Psi}_0(k,q,x)$ as functions of $x\equiv k\tau$ with two different
scales ($k/C=10$ and $100$) and two different momenta ($m/q=10$ and $100$).
Both $\tilde{\Psi}_0(k,q,x)$ and $\tilde{\Psi}_2(k,q,x)$ are calculated
from the exact solution (solid line), or fluid approximation
with $l_{\rm max}=2$ (dotted line) and $3$ (dashed line).
}%
\label{psi2ovpsi0}
\end{figure*}
%%%%%%%%%%%%%%%%%%%%%%%%%%%%%%%%%%%%%%%%%%%%%%%%%%%%%%%%%%%%%%%%%%
\subsubsection{$\tilde{\Psi}_0(k,q,x)$}
Figure \ref{psi0} shows the evolution of $\tilde{\Psi}_0(k,q,x)$
for two different scales ($k/C=10$ and $100$) and two different momenta
($m/q=10$ and $100$). We calculate $\tilde{\Psi}_0^{\rm fluid}(k,q,x)$
by truncating the Boltzmann equations at $l_{\rm max}=1$, $2$ and $3$
(fluid approximation), while we calculate $\tilde{\Psi}_0^{\rm exact}(k,q,x)$
from the exact solution of the Boltzmann equations given in Eq.(\ref{eq:psi0_analytic}).
The fractional error in the fluid approximation is defined as
$\Delta\tilde{\Psi}_0/\tilde{\Psi}_0\equiv\tilde{\Psi}_0^{\rm fluid}/\tilde{\Psi}_{0}^{\rm exact}-1$.
For each combination of $m/q$ and $k/C$, we show both
$\tilde{\Psi}_0^{\rm exact}(k,q,x)$ (top),
and $|\Delta\tilde{\Psi}_0/\tilde{\Psi}_0|$ (bottom) with
$l_{\rm max}=1$ (solid lines), $2$ (dotted lines) and $3$ (dashed lines).

At large-scale ($k/C=10$ or $k\simeq 0.1~h~{\rm Mpc^{-1}}$),
neutrinos with $m/q=10$ become non-relativistic at around the horizon re-entry,
and neutrinos with $m/q=100$ become non-relativistic well {\it before} the horizon re-entry
($x_{nr}=k\tau_{nr}=0.98$ and $0.10$ for $m/q=10$ and $100$, respectively),
and fractional errors of the fluid approximation peak at $x=x_H\sim 1$.
At small-scale ($k/C=100$ or $k\simeq 1.0~h~{\rm Mpc^{-1}}$),
neutrinos with $m/q=100$ become non-relativistic at around the horizon re-entry, while
neutrinos with $m/q=10$ become non-relativistic well {\it after} the horizon re-entry
($x_{nr}=k\tau_{nr}=8.3$ and $0.98$ for $m/q=10$ and $100$, respectively).
For neutrinos becoming non-relativistic after the horizon re-entry,
fractional errors of the fluid approximation peak at $x=x_{nr}$.

We see that the asymptotic growth rate is
$\tilde{\Psi}_0^{\rm exact}(k,q,x)\propto x^2\propto a$
(see Eq.(\ref{eq:psi0_asymp})).
For neutrinos with $m/q=10$ and $k/C=100$, the growth of
$\tilde{\Psi}_0^{\rm exact}(k,q,x)$
is suppressed between the horizon re-entry and the epoch of relativistic to
non-relativistic transition ($x_{nr}=8.3$); however, once neutrinos become non-relativistic,
$\tilde{\Psi}_0^{\rm exact}(k,q,x)$ grows rapidly, catching up with the gravitational
potential set up by CDM.
\footnote{
Even though we do not include CDM explicitly, by setting $\dot{\phi}=\dot{\psi}=0$,
we are including CDM as a dominant source of the gravitational potential in the
Boltzmann equations (i.e., the last term of Eq.(\ref{eq:psi1})).
If we calculate $\psi(k,x)$ including the suppression of the gravitational potential
due to massive neutrino free-streaming, the neutrino catch-up will be slightly slower.
}
For both large and small-scales, fluid approximation becomes more accurate
as we increase the $l_{\rm max}$, but for neutrinos with $m/q=100$ and $k/C=10$,
$l_{\rm max}=1$ is sufficient to approximate the exact solution to better than $1\%$
accuracy for almost entire evolution history ($x_H<x<x_0$).
For small-scale, $k/C=100$, with nearly relativistic neutrinos ($m/q=10$),
fluid approximation with low multipole cutoff (i.e., $l_{\rm max}=1$, $2$ and $3$)
breaks down for almost entire evolution history, while at late time ($x>1000$),
we start to see an accuracy of better than $1\%$ for $l_{\rm max}=2$ and $3$.

\subsubsection{$\tilde{\Psi}_1(k,q,x)$}
Figure \ref{psi1} shows the evolution of $\tilde{\Psi}_1(k,q,x)$
for two different scales ($k/C=10$ and $100$) and two different momenta
($m/q=10$ and $100$).
We calculate $\tilde{\Psi}_1^{\rm fluid}(k,q,x)$ and the fractional error in the fluid
approximation in the same ways as before, while we calculate
$\tilde{\Psi}_1^{\rm exact}(k,q,x)$ from the exact solution of the Boltzmann
equations given in Eq.(\ref{eq:psi1_analytic}).
Results are almost the same as the case for $\tilde{\Psi}_0(k,q,x)$, except
for the asymptotic growth rate is
$\tilde{\Psi}_1^{\rm exact}(k,q,x)\propto x^3\propto a^{3/2}$
(see Eq.(\ref{eq:psi1_asymp})).
Since $\delta(k,x)\propto\tilde{\Psi}_0(k,q,x)\propto x^2$
and $\theta(k,x)\propto\tilde{\Psi}_1(k,q,x)\propto x^3$,
the late-time evolution of $\delta(k,x)$ and $\theta(k,x)$
from the Boltzmann equations is consistent with the continuity equation
(i.e., $\dot{\delta}(k,\tau)=-\theta(k,\tau)$).

\subsubsection{$\tilde{\Psi}_2(k,q,x)$}
Figure \ref{psi2ovpsi0} shows the ratio of $\tilde{\Psi}_2(k,q,x)$
to $\tilde{\Psi}_0(k,q,x)$
for exact solutions (solid lines), $l_{\rm max}=2$ (dotted lines) and
$l_{\rm max}=3$ (dashed lines) for two different scales ($k/C=10$ and $100$)
and two different momenta ($m/q=10$ and $100$).
As we discussed in \S~\ref{sec:boltzmann_hierarchy}, we have a useful
relation between $\tilde{\Psi}_0(k,q,x)$ and $\tilde{\Psi}_2(k,q,x)$
if we truncate the Boltzmann equations at $l_{\rm max}=2$ (see Eq.(\ref{eq:ansatz})).
Here, we check how late-time evolution can simplify the relation
between $\tilde{\Psi}_0(k,q,x)$ and $\tilde{\Psi}_2(k,q,x)$.
At sufficiently late time, we have an asymptotic value of
$\tilde{\Psi}_2(k,q,x)=-\frac25\tilde{\Psi}_0(k,q,x)$, which allows us
to replace the anisotropic stress term in the fluid equations by the
sound speed (Eq.(\ref{eq:sigtopressure})).
As we see in the figure, the asymptotic value of $\tilde{\Psi}_2(k,q,x)/\tilde{\Psi}_0(k,q,x)=-2/5$ is reached at relatively early times.

\subsubsection{Summary}
We found that the applicability of the fluid approximation
on the perturbed distribution function, $\tilde{\Psi}_l(k,q,x)$,
depends crucially on our choice of wavenumber, $k$, momentum,
$q$ and mass, $m$, of particles.
Generally speaking, wavenumber, $k$, sets the time of the horizon crossing,
$a_H=a(x\sim1)$, and the momentum and mass of the
particles give the epoch of the relativistic to non-relativistic
transition, $a_{\rm nr}\simeq q/m$. As long as a given particle becomes
non-relativistic before the horizon crossing, $a_{\rm nr}< a_H$,
the fluid approximation and the exact solution of $\tilde{\Psi}_l(k,q,x)$
agree to better than $1\%$ accuracy.
Re-writing $a_{\rm nr}<a_H$ in terms of $m/q$ and $k/C$, we have,
\begin{eqnarray}
\frac{q}{m}<\frac{C}{k}+\frac14\left(\frac{C}{k}\right)^2.
\end{eqnarray}
This condition can be easily satisfied for a large-scale mode,
where $C/k\gg 1$ ($k\ll 0.01~h~{\rm Mpc^{-1}}$), and the condition can be
satisfied in a small-scale mode, $C/k\ll 1$ ($k\gg 0.01~h~{\rm Mpc^{-1}}$),
if $\frac{m}{q}>\frac{k}{C}$.

So far, we discussed the validity of the fluid approximation for
$\tilde{\Psi}_l(k,q,x)$ with a given momentum and a given wavenumber.
However, quantities such as $\delta(k,x)$ and $\theta(k,x)$ are given
as integrals of $\tilde{\Psi}_l(k,q,x)$ over momentum space ($0<q<\infty$)
weighted by $f_0(q)$ and appropriate powers of $q$.
The unperturbed distribution function, $f_0(q)$, exponentially suppresses
the population of nearly relativistic neutrinos with an exponent of
$\frac{q}{m_{\rm\nu}}\frac{m_{\rm\nu}}{T_{\rm\nu,0}}\frac{a_{eq}}{a_0}$ (see Eq.(\ref{eq:f0_schematic})).
For a given lower limit of neutrino mass, $m_{\rm\nu}>0.05~{\rm eV}$, and 
the current temperature of neutrinos, $T_{\rm\nu,0}\sim 1.9~{\rm K}$,
we have $\frac{m_{\rm\nu}}{T_{\rm\nu,0}}>340$.
Therefore, the population of relativistic neutrinos with $\frac{m}{q}\ll0.1$
is negligible, and does not affect calculations of $\delta(k,x)$ or $\theta(k,x)$.

\subsection{Limitation of Fluid Approximation on $\delta_{\rm\nu}(k,x)$}
\label{sec:fluid_approx_neutrino}
Here, we study the limitation of the fluid approximation on the
neutrino density contrast, $\delta_{\rm\nu}(k,x)$, and
the velocity divergence, $\theta_{\rm\nu}(k,x)$, with two different
masses of neutrinos, $m_{\rm\nu}=0.05$ and $0.5~{\rm eV}$.
%\footnote{%
%Even though our study focuses on the most massive neutrino flavor
%with $0.05\le m_{\rm\nu,i}\le 0.58~{\rm eV}$, we also consider the
%case with $m_{\rm\nu,i}=0.01~{\rm eV}$ as an example of close-to-massless neutrinos.
%}%
As we have seen in the previous section, for given $k$ and $a_0/a_{eq}$, the ratio of the
neutrino mass to the current temperature of neutrinos determines the validity
of the fluid approximation.

With the current temperature of neutrinos, $T_{\rm\nu,0}\sim 1.9~{\rm K}$,
we have $m_{\rm\nu}/T_{\rm\nu,eq}=305$ and $3050$ for $m_{\rm\nu}=0.05$
and $0.5~{\rm eV}$, respectively.
To find $\delta_{\rm\nu}(k,x)$ and $\theta_{\rm\nu}(k,x)$, we integrate
$\tilde{\Psi}_l(k,q,x)$ using Eqs.(\ref{eq:rho_bar})$\sim$(\ref{eq:sigma})
for several $l_{\rm max}$.
%%%%%%%%%%%%%%%%%%%%%%%%%%%%%%%%%%%%%%%%%%%%%%%%%%%%%%%%%%%%%%%%%%
%insert figure here
\begin{figure*}[t]
\rotatebox{0}{%
  \includegraphics[width=16cm]{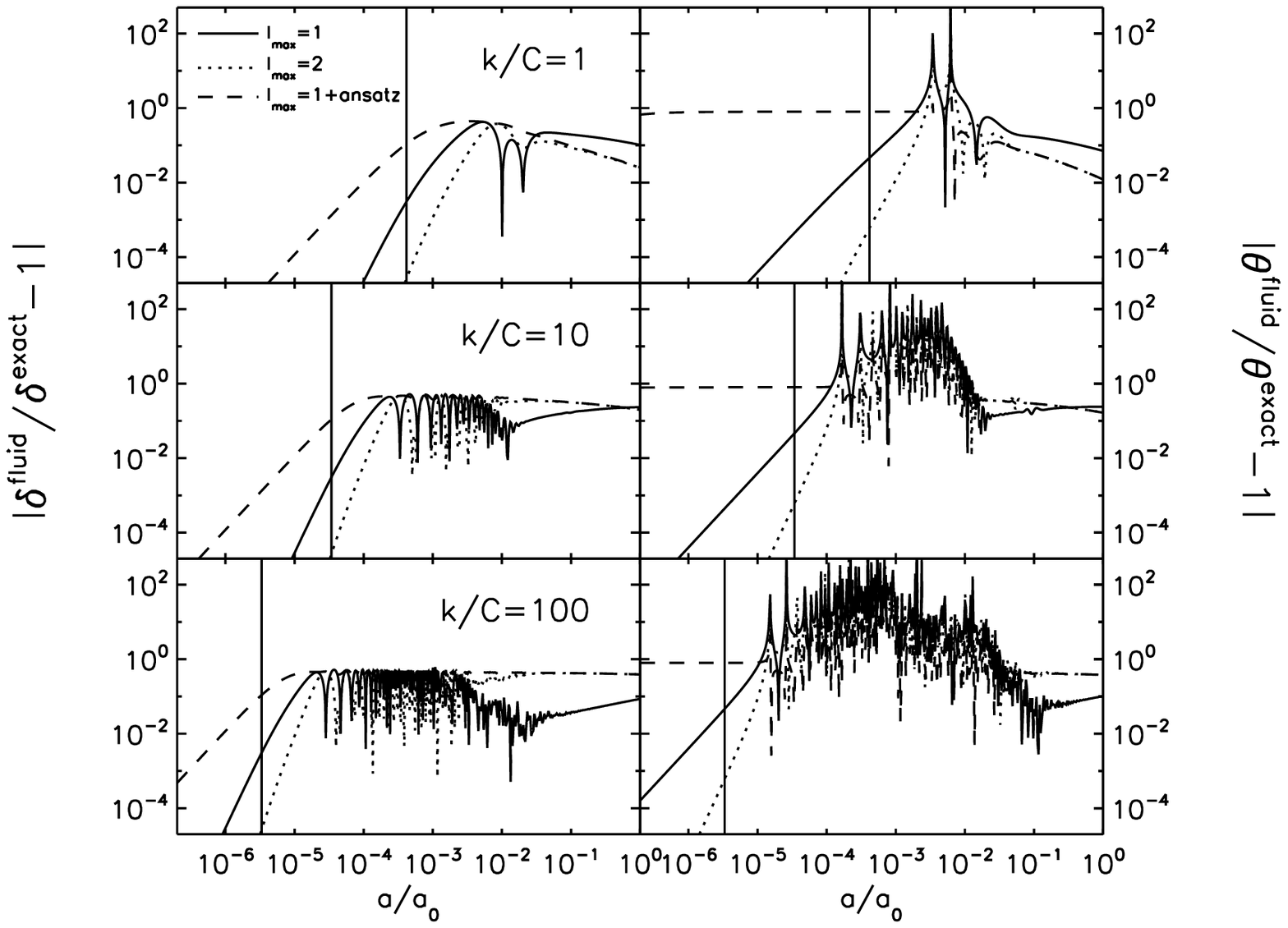}
}%
\caption{%
({\it left}):
Time evolution of the fractional errors of 
$\Delta\delta_{\rm\nu}(k,x)/\delta_{\rm\nu}(k,x)\equiv\delta_{\rm\nu}^{\rm fluid}(k,x)/
\delta_{\rm\nu}^{\rm exact}(k,x)-1$.
The solid lines show $l_{\rm max}=1$, while the dotted lines show $l_{\rm max}=2$.
The dashed lines show $l_{\rm max}=1$, but with the ansatz
for $l=2$, $\tilde{\Psi}_2(k,q,x)=-\frac25\tilde{\Psi}_0(k,q,x)$.
({\it right}):
Time evolution of the fractional errors of 
$\Delta\theta_{\rm\nu}(k,x)/\theta_{\rm\nu}(k,x)\equiv\theta_{\rm\nu}^{\rm fluid}(k,x)/
\theta_{\rm\nu}^{\rm exact}(k,x)-1$.
Here, we use $m_{\rm\nu}=0.05~{\rm eV}$, and show the results at three
different scales, $k/C=1, 10$ and $100$, corresponding to
$k\simeq0.01, 0.1$ and $1.0~h~{\rm Mpc^{-1}}$, respectively.
The vertical lines show the time of the horizon crossing for
each mode. The present-day scale factor is $a_0=3000$.
}%
\label{mn=0.001}
\end{figure*}
%%%%%%%%%%%%%%%%%%%%%%%%%%%%%%%%%%%%%%%%%%%%%%%%%%%%%%%%%%%%%%%%%%
%insert figure here
\begin{figure*}[t]
\rotatebox{0}{%
  \includegraphics[width=16cm]{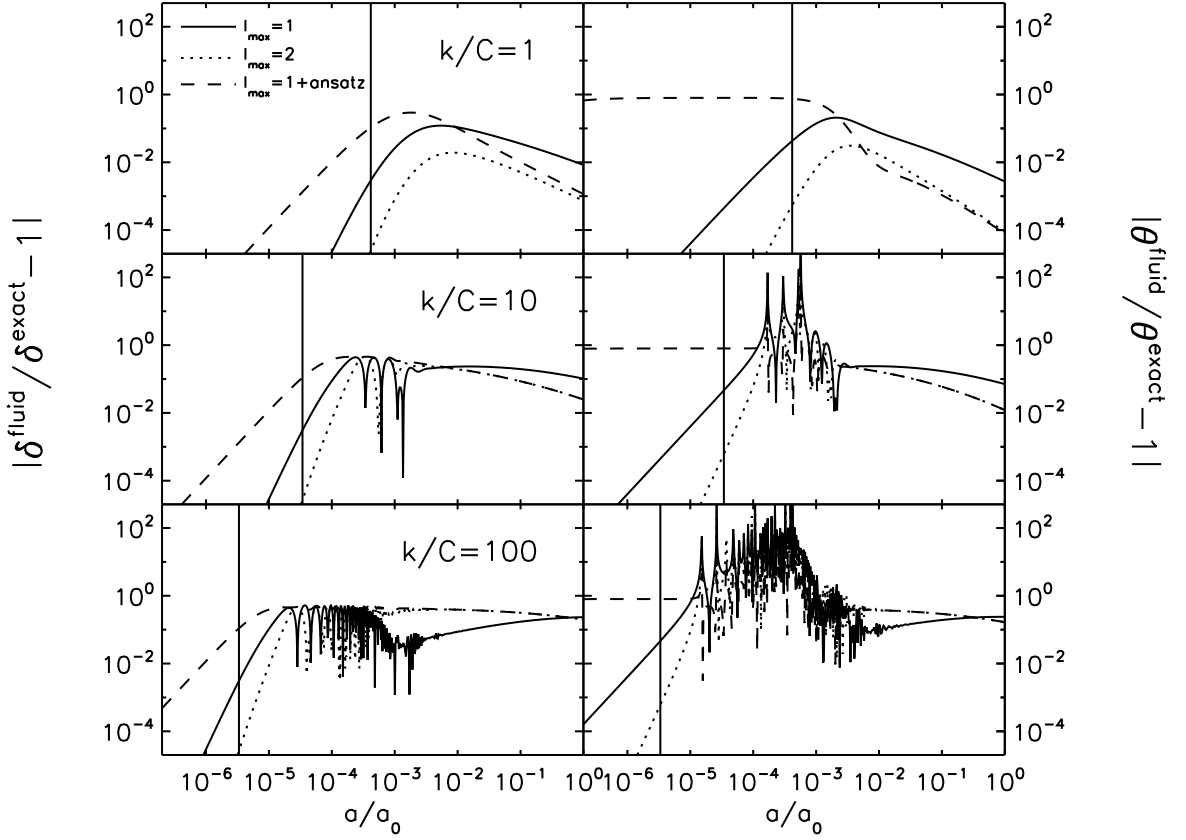}
}%
\caption{%
Same as Figure \ref{mn=0.001} for $m_{\rm\nu}=0.5~{\rm eV}$.
}%
\label{mn=0.1}
\end{figure*}
%%%%%%%%%%%%%%%%%%%%%%%%%%%%%%%%%%%%%%%%%%%%%%%%%%%%%%%%%%%%%%%%%%
\subsubsection{$\delta_{\rm\nu}(k,x)$ and $\theta_{\rm\nu}(k,x)$ with small $m_{\rm\nu}$}
\label{sec:mn=0.001}
Figure \ref{mn=0.001} shows the evolution of the fractional errors of the fluid approximation,
$\Delta\delta_{\rm\nu}(k,x)/\delta_{\rm\nu}(k,x)\equiv\delta_{\rm\nu}^{\rm fluid}(k,x)/
\delta_{\rm\nu}^{\rm exact}(k,x)-1$ and $\Delta\theta_{\rm\nu}(k,x)/\theta_{\rm\nu}(k,x)
\equiv\theta_{\rm\nu}^{\rm fluid}(k,x)/\theta_{\rm\nu}^{\rm exact}(k,x)-1$,
as functions of $a/a_0$ for three different
scales $k/C=1$, $10$ and $100$ with $m_{\rm\nu}=0.05~{\rm eV}$
($C=0.01~h~{\rm Mpc^{-1}}$).
As expected, the fluid approximation does not yield accurate results for
such a small mass.
As was the case for $\tilde{\Psi}_l^{\rm fluid}(k,q,x)$, the fractional error
increases shortly after the horizon entrance, and then decreases as neutrinos
become non-relativistic.
For neutrinos with $m_{\rm\nu}=0.05~{\rm eV}$, Eq.(\ref{eq:znr}) gives
$a_{\rm nr}/a_0=0.01$, or $x_{nr}=9.4\frac{k}{C}$,
and as a result the fluid approximation breaks down during entire
evolution history of $\delta_{\rm\nu}(k,x)$ and $\theta_{\rm\nu}(k,x)$ between
the horizon crossing and the present epoch ($a_H<a<a_0$).
Nevertheless, for the largest scale ($k/C=1$),
the error is below 10\% level at low redshit, as the neutrinos become
sufficiently non-relativistic.
We also show the fractional errors of fluid approximation for
$\delta_{\rm\nu}(k,x)$ and $\theta_{\rm\nu}(k,x)$ calculated
using the late time asymptotic value of Eq.(\ref{eq:ansatz}):
$\tilde{\Psi}_2(k,q,x)=-\frac25\tilde{\Psi}_0(k,q,x)$ (dashed lines).
As we have seen in Figure \ref{psi2ovpsi0}, this simple ansatz works well
and follows the fractional error with $l_{\rm max}=2$ at late time, $a\ll a_{\rm nr}$.

\subsubsection{$\delta_{\rm\nu}(k,x)$ and $\theta_{\rm\nu}(k,x)$ with large $m_{\rm\nu}$}
\label{sec:mn=0.1}
Figure \ref{mn=0.1} shows the evolution of the fractional errors of a fluid
approximation as functions of $a/a_0$ for three different
scales $k/C=1$, $10$ and $100$ with $m_{\rm\nu}=0.5~{\rm eV}$.
For neutrinos with $m_{\rm\nu}=0.5~{\rm eV}$, Eq.(\ref{eq:znr}) gives
$a_{\rm nr}/a_0=0.001$, or $x_{nr}=2.1\frac{k}{C}$,
and all the scale with $k/C\gtrsim 0.5$ enter the horizon
when neutrinos are relativistic. As a result, the fluid
approximation with $l_{\rm max}=1$ is only accurate to $\sim1\%$ at large scale
($k/C=1$) and $\sim20\%$ at small scale ($k/C=100$) at low redshift.

We see that $l_{\rm max}=2$ and the ansatz approximates the small-scale density
contrast and velocity divergence better than the case with $l_{\rm max}=1$,
and the fluid approximation becomes accurate to $\lesssim1\%$
at large scale ($k/C=1$).

\subsection{Range of Validity of Fluid Approximation}
\label{sec:range_of_FA}
\begin{figure*}[t]
\rotatebox{0}{%
  \includegraphics[width=16cm]{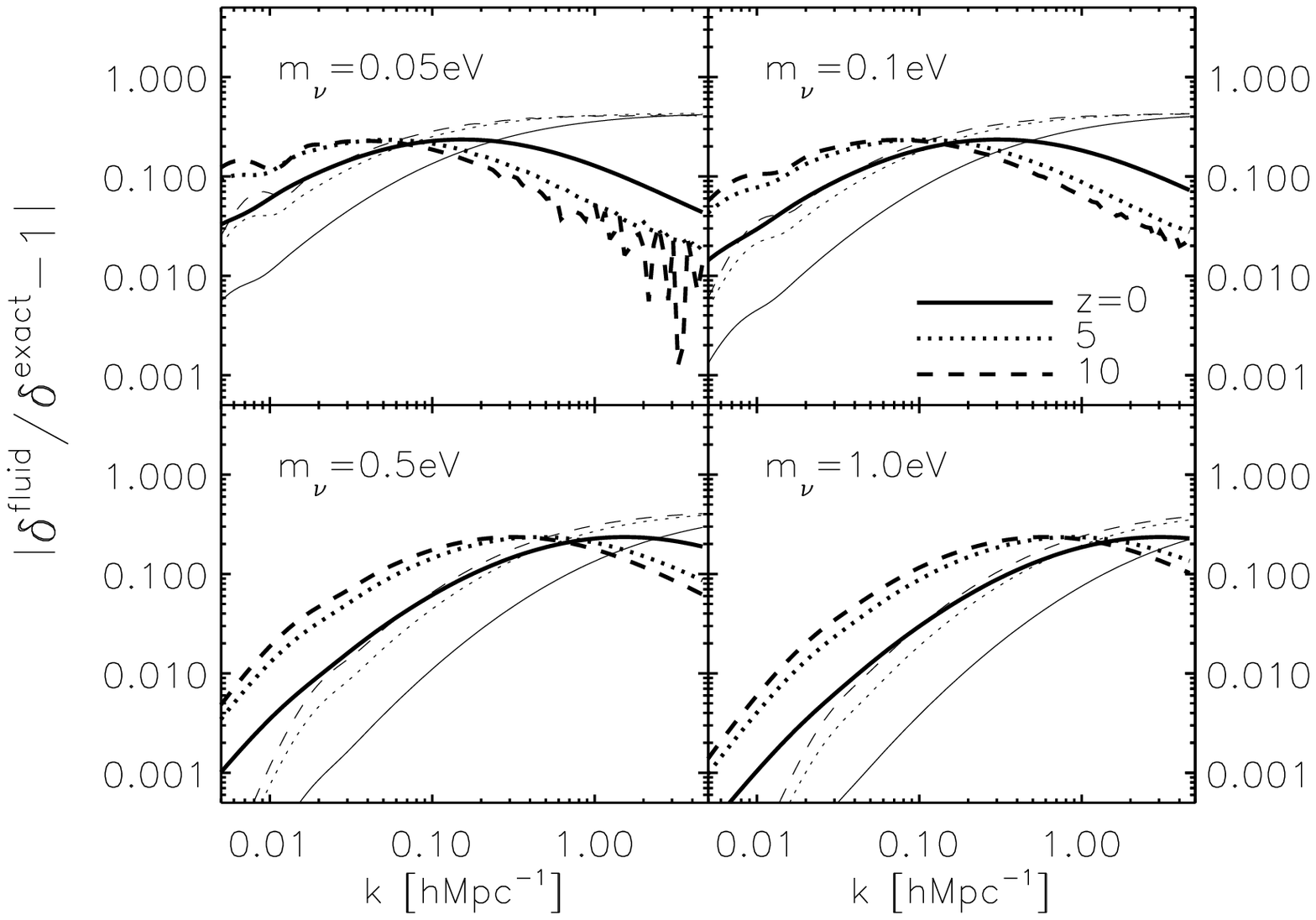}
}%
\caption{%
We show the fractional errors,
$\Delta\delta_{\rm\nu}(k,x)/\delta_{\rm\nu}(k,x)\equiv\delta_{\rm\nu}^{\rm fluid}(k,x)/
\delta_{\rm\nu}^{\rm exact}(k,x)-1$, for four different masses of neutrino
at three different redshifts, $z=0,~5$ and $10$ as functions of
wavenumber, where the thick and thin lines are for $l_{\rm max}=1$ and $2$, respectively.
}%
\label{kmax}
\end{figure*}

We have seen that the fractional errors of fluid approximation for $\delta_{\rm\nu}(k,x)$ and
$\theta_{\rm\nu}(k,x)$ decrease for heavier particles, but the errors are still significant
on small-scales.
Now, the question is what is the maximum wavenumber, $k_{max}$, below which we can use
the fluid approximation with 10 or 20\% accuracy for a given mass of neutrinos
at a given time.
Figure \ref{kmax} shows the fractional error,
$\Delta\delta_{\rm\nu}(k,x)/\delta_{\rm\nu}(k,x)\equiv\delta_{\rm\nu}^{\rm fluid}(k,x)/
\delta_{\rm\nu}^{\rm exact}(k,x)-1$, for four different masses of neutrinos
at three different redshifts, $z=0, 5$ and $10$.
We find that the fluid approximation is only accurate to
few$\sim25\%$ over a wide range of $k$ at low redshift.

Table \ref{tb:kmax} shows $k_{max}$ with $l_{\rm max}=1$
and $2$ for various neutrino masses,
$m_{\rm\nu}=0.05,~0.10,~0.50$ and $1.0~{\rm eV}$, at five redshifts,
$z=0,~1,~3,~5$ and $10$.
The smaller the redshift is and the larger $m_{\rm\nu}$ is,
the larger $k_{max}$ becomes.
We see that $k_{max}$ is $3\sim4$ times larger with $l_{\rm max}=2$
than with $l_{\rm max}=1$.

We are particularly interested in the $k$-range of linear to mildly
non-linear regime on the matter density power spectrum at the
redshifts relevant to the future and on-going galaxy redshift surveys
($z\lesssim3$),
and for that purpose, $0.1\lesssim k_{max}\lesssim 0.4~h~{\rm Mpc^{-1}}$
will be necessary with sufficient accuracy: on smaller scales,
the non-linearity is too large for power spectrum to be
used for cosmology.
%%%%%%%%%%%%%%%%%%%%%%%%%%%%%%%%%%%%%%%%%%%%%%%%%%%%%%%%%%%%%%%%%%
%insert figure here
%%%%%%%%%%%%%%%%%%%%%%%%%%%%%%%%%%%%%%%%%%%%%%%%%%%%%%%%%%%%%%%%%%
\begin{table*}
\begin{tabular}{|l|c|c|c|c|}
\hline
$l_{\rm max}=1$ & $m_{\rm\nu}=0.05~{\rm eV}$ & $0.1~{\rm eV}$ & $0.5~{\rm eV}$ & $1.0~{\rm eV}$ \\
\hline
z=0      &  0.009 (0.032) & 0.018 (0.064) & 0.090 (0.31) & 0.18 (0.62)\\
z=1      & 0.007 (0.021) & 0.012 (0.042) & 0.060 (0.19) & 0.12 (0.38)\\
z=3      & 0.005 (0.015) & 0.009 (0.028) & 0.039 (0.13) & 0.079 (0.25)\\
z=5      & 0.003 (0.012) & 0.008 (0.023) & 0.032 (0.097) & 0.060 (0.19)\\
z=10     & 0.002 (0.009) & 0.004 (0.017) & 0.023 (0.068) & 0.042 (0.14)\\
\hline
\hline
$l_{\rm max}=2$ & $m_{\rm\nu}=0.05~{\rm eV}$ & $0.1~{\rm eV}$ & $0.5~{\rm eV}$ & $1.0~{\rm eV}$ \\
\hline
z=0      & 0.037 (0.097) & 0.073 (0.19) & 0.36 (0.94) & 0.72 (1.88)\\
z=1      & 0.024 (0.064) & 0.045 (0.13) & 0.22 (0.58) & 0.44 (1.16)\\
z=3      & 0.016 (0.042) & 0.032 (0.079) & 0.15 (0.36) & 0.27 (0.72)\\
z=5      & 0.013 (0.034) & 0.024 (0.064) & 0.11 (0.29) & 0.22 (0.54)\\
z=10     & 0.009 (0.024) & 0.018 (0.045) & 0.079 (0.19) & 0.15 (0.38)\\
\hline
\end{tabular}
\caption{%
The maximum wavenumber, $k_{max}[h~{\rm Mpc^{-1}}]$,
for which the fluid approximation is accurate at 10 (20)\% or better.
}%
\label{tb:kmax}
\end{table*}
%%%%%%%%%%%%%%%%%%%%%%%%%%%%%%%%%%%%%%%%%%%%%%%%%%%%%%%%%%%%%%%%%%
\section{Discussions and Conclusions}
\label{sec:conclusions}
We have calculated the evolution of the perturbed distribution
functions of massive neutrinos, $\tilde{\Psi}_l(k,q,x)$, using the fluid approximation,
i.e., truncation of the Boltzmann equations at $l_{\rm max}=1$, $2$ and $3$.
We compared the approximate solutions to the exact solutions that we
have derived in this paper.
When the distribution function is dominated by the relativistic neutrinos,
fluid approximation poorly represents the exact oscillation phase of
$\tilde{\Psi}_l(k,q,x)$ calculated from the exact solution.
When the distribution function is dominated by non-relativistic neutrinos,
$\tilde{\Psi}_0(k,q,x)$ and $\tilde{\Psi}_1(k,q,x)$ are sourced mainly by the
gravitational potential, $\psi(k,x)$, and decoupled from the higher multipoles,
$\tilde{\Psi}_{l\ge2}(k,q,x)$.
This allows the fluid approximation to be an excellent approximation to the growth of
$\tilde{\Psi}_0(k,q,x)$ and $\tilde{\Psi}_1(k,q,x)$ for small $q$.
Then, we integrated the perturbed distribution functions
to calculate the quantities such as $\delta_{\rm\nu}(k,x)$ and
$\theta_{\rm\nu}(k,x)$.
Comparing the density contrasts of massive neutrinos calculated from
the fluid approximation to the exact solutions,
we found that the fluid approximation is only accurate to few$\sim25\%$
for $k\lesssim0.4~h~{\rm Mpc^{-1}}$ and $0.05\le m_{\rm\nu}\le0.5{\rm eV}$
To increase the accuracy of the fluid approximation further, it is
necessary to either directly solve for the Boltzmann hierarchy
with $l_{max}\ge3$ as in Eqs.(\ref{eq:psi0})$\sim$(\ref{eq:psil}), or
solve fluid equations, Eqs.(\ref{eq:continuity}) and (\ref{eq:Euler}),
with an ansatz for an anisotropic stress,
$k^2\sigma(k,\tau)$, as we did for $l_{max}=2$ in Eq.(\ref{eq:sigtopressure}).

We solved the Boltzmann equation for massive neutrinos in an EdS universe for which
$\dot{\phi}(k,x)=\dot{\psi}(k,x)=0$. In a more realistic multi-component fluid
case, we have $\dot{\phi}(k,x)\ne 0$ and $\dot{\psi}(k,x)\ne 0$ due to
the effect of massive neutrinos even during the matter dominated
epoch. Including this effect is straightforward.
For the observationally allowed range of neutrino masses, we expect the correction
to be small, as the dominant source of gravitational potential is
still CDM ($f_{\rm\nu}\lesssim 0.05$).
Therefore, our conclusions regarding the limitation of the fluid approximation is
not affected by our using an EdS universe.
During the dark energy dominated epoch, the EdS approximation
breaks down, and $\phi$ and $\psi$ evolve.
Nevertheless, the correction will be limited to the dark energy dominated epoch,
$a>a_{DE}$, and the scale around $k\gtrsim k_{\rm FS}(a_{DE})\gg k_{\rm FS}(a_{\rm nr})$.
We found that, as long as the term proportional to the gravitational potential,
$\psi(k,x)$, dominates the right hand side of Eq.(\ref{eq:psi1}),
the fluid approximation is valid. Therefore, unless the effect of the dark energy
suppresses $\psi(k,x)$ much faster than the growth of $\epsilon^2(q,x)\propto a^2$,
$k_{max}$ at $a>a_{DE}$ will not change significantly.

Since we have studied the evolution of the distribution function solving
the collision-less Boltzmann equation, one can apply these results to other
collision-less particles in general.

Now, why is fluid approximation useful?
Future and on-going dark energy missions aim at the accurate measurement
of the galaxy/matter power spectrum with an accuracy better than 1\%.
One might think that the cosmological linear perturbation theory has
already been well established, and the numerical codes such as CMBfast
and CAMB can calculate the linear matter power spectrum with an accuracy
better than 1\%.

However, the linear perturbation theory breaks down at small-scale and
low redshift, where the density contrast becomes non-linear
($k\gtrsim0.1~h~{\rm Mpc^{-1}}$ at $z\sim1$)
\cite{jeong/komatsu:2006,carlson/white/padmanabhan:2009}.
Therefore, in order to exploit the cosmological information contained in
a given survey, one needs to understand the non-linearities on the
galaxy/matter power spectrum
\cite{yamamoto/bassett/nishioka:2005,rassat/etal:2008,shoji/jeong/komatsu:2009}.

Among the non-linearities, the matter clustering has been well understood in
the mildly non-linear regime
(see \cite{bernardeau/etal:2002}, for a review), but the theories have been limited
to CDM dominated universe. The pressure gradient term in the Euler
equation was completely ignored.

In our previous work, we developed the 3rd-order perturbation theory
with the pressure gradient terms explicitly included \cite{shoji/komatsu:2009}
(also see \cite{saito/takada/taruya:2008,wong:2008,lesgourgues/etal:2009,mcdonald:2009}).
With this extension to the higher order perturbation theory
as well as within the limitation on the accuracy of $\delta_{\rm \nu}$
calculated from the fluid approximation, we can
now calculate the next-to-linear order matter power spectrum with
massive neutrino free-streaming effect, properly included.

Since the structure formation is mostly affected by the most massive species
of neutrinos, and the current constraints on the total mass of neutrinos indicate
that at least one of the neutrino species has a mass of order a tenth of
${\rm eV}$, the use of fluid approximation is limited with an accuracy of
few to 25\% over $k\lesssim0.4~h~{\rm Mpc^{-1}}$ for $z<10$.
As a result, for a small fraction of massive neutrino,
$f_{\rm\nu}\lesssim0.04$ for $\sum_im_{\rm\nu,i}\lesssim0.5~{\rm eV}$,
the fractional error on the matter density contrast,
$\delta_m=(1-f_{\rm\nu})\delta_{\rm c}+f_{\rm\nu}\delta_{\rm\nu}$,
calculated with the fluid approximation is accurate to sub-percent level.

%%%%%%%%%%%%%%%%%%%%%%%%%%%%%%%%%%%%%%%%%%%%%%%%%%%%%%%%%%%%%%%%%%%
This material is based in part upon work 
supported by the Texas Advanced Research Program under 
Grant No. 003658-0005-2006, by NASA grants NNX08AM29G 
and NNX08AL43G, and by NSF grant AST-0807649.
M.~S. thanks for warm hospitality of Astronomical Institute at
Tohoku University where part of this work was done.
%%%%%%%%%%%%%%%%%%%%%%%%%%%%%%%%%%%%%%%%%%%%%%%%%%%%%%%%%%%%%%%%%%
\appendix
\begin{widetext}
%%%%%%%%%%%%%%%%%%%%%%%%%%%%%%%%%%%%%%%%%%%%%%%%%%%%%%%%%%%%%%%%%%
\section{Sound speed versus Velocity Dispersion}
\label{sec:sigma_app}
The wavenumber corresponding to the free-streaming scale, $k_{\rm FS}$,
is defined by the single-fluid continuity and Euler equations:
\begin{eqnarray}
&&\dot{\delta}(\mathbf{k},\tau)+\theta(\mathbf{k},\tau)=0\\
&&\dot{\theta}(\mathbf{k},\tau)+\mathcal{H}(\tau)\theta(\mathbf{k},\tau)
+c_{\rm s}^2(\tau)\left[k_{\rm FS}^2(\tau)-k^2(\tau)\right]\delta(\mathbf{k},\tau)=0,
\nonumber\\
\end{eqnarray}
where
\begin{eqnarray}
k_{\rm FS}(\tau)\equiv\sqrt{\frac32}\frac{\mathcal{H}(\tau)}{c_{\rm s}(\tau)},
\end{eqnarray}
is the scale that divides the characteristics of the solution, $\delta(\mathbf{k},\tau)$.
For $k<k_{\rm FS}$, $\delta(\mathbf{k},\tau)$ grows, while for
$k>k_{\rm FS}$, $\delta(\mathbf{k},\tau)$ oscillates.

In the literature, however, the sound speed, $c_{\rm s}(\tau)$, has often been
replaced, or crudely approximated, by the velocity dispersion, $\sigma_{\rm\nu}$,
such that $c_{\rm s}\simeq\sigma_{\rm\nu,i}$ without any justification.
Strictly speaking, the velocity dispersion should not be used to define
the free-streaming scale, $k_{\rm FS}$, as the Euler equation contains
sound horizon, $c_{\rm s}^2\equiv\frac{\delta P}{\delta\rho}$,
not the velocity dispersion.
In this Appendix, we shall clarify this issue.

The sound speed is given by
\begin{eqnarray}
c_s^2(k,\tau)\equiv\frac{\delta P(k,\tau)}{\delta\rho(k,\tau)}
=\frac13 \frac{\int q^2dq\frac{q^2}{\epsilon(q,\tau)}f_0(q)\Psi_0(k,q,\tau)}
{\int q^2dq\epsilon(q,\tau)f_0(q)\Psi_0(k,q,\tau)},
\label{eq:cs_app}
\end{eqnarray}
and the velocity dispersion is given by
\begin{eqnarray}
\sigma_{\rm\nu}^2(\tau)\equiv
\frac{\int q^2dq\left[\frac{q}{\epsilon(q,\tau)}\right]^2f_0(q)}
{\int q^2dqf_0(q)}.
\label{eq:sig_app}
\end{eqnarray}
Note that the sound speed can depend on $k$ in general, while velocity dispersion is,
by definition, independent of $k$.

As we see from Eq.(\ref{eq:psi0_asymp}),
in the non-relativistic limit, where the phase-space distribution of the
particles is dominated by non-relativistic particles ($q\ll a(x)m$),
the $k$ dependence of
$\Psi_0(k,q,\tau)\equiv\tilde{\Psi}_0(k,\tau)\frac{d\ln f_0(q)}{d\ln q}$
will be canceled both in the denominator and numerator of Eq.(\ref{eq:cs_app}).
If perturbations are adiabatic, i.e.,
$\frac{\delta P}{\delta\rho}=\frac{\dot{\bar{P}}(\tau)}{\dot{\bar{\rho}}(\tau)}$,
we have
\begin{eqnarray}
c_s^2(\tau)=\frac{\dot{\bar{P}}(\tau)}{\dot{\bar{\rho}}(\tau)}
=w(\tau)-\frac{\dot{w}(\tau)}{3\mathcal{H}(\tau)[1+w(\tau)]},
\end{eqnarray}
where $w(\tau)\equiv\frac{\bar{P}(\tau)}{\bar{\rho}(\tau)}$
is an equation of state.
Since the velocity dispersion {\it in the non-relativistic} limit is given as
\begin{eqnarray}
\sigma_{\rm\nu}^2(\tau)\to\frac1{a^2(\tau)m^2}
\frac{\int q^2dq~q^2f_0(q)\Psi_0(k,q,\tau)}
{\int q^2dq f_0(q)\Psi_0(k,q,\tau)}=3w(\tau),
\end{eqnarray}
we have
\begin{eqnarray}
c_s^2(\tau)\to\frac13\sigma_{\rm\nu}^2(\tau)
+\frac29\frac{\sigma_{\rm\nu}^2(\tau)}{1+\frac13\sigma_{\rm\nu}^2(\tau)}
\simeq\frac59\sigma_{\rm\nu}^2(\tau).
\end{eqnarray}
Here, we have used $\sigma_{\rm\nu}^2(\tau)\ll 1$.
Therefore, in the non-relativistic limit, we have
$c_s=\frac{\sqrt{5}}{3}\sigma_{\rm\nu,i}\simeq 0.745\sigma_{\rm\nu,i}$.
%%%%%%%%%%%%%%%%%%%%%%%%%%%%%%%%%%%%%%%%%%%%%%%%%%%%%%%%%%%%%%%%%%
\section{Exact Analytic Solution of $\tilde{\Psi}_l(k,q,x)$}
\label{sec:exact_solution_app}
We derive the analytic solutions of the Boltzmann equation.
We start from Eq.(\ref{eq:boltzmann2}),
\begin{eqnarray}
\frac{\partial\tilde{\Psi}(k,q,\mu,x)}{\partial x}+i\frac{q}{\epsilon(q,x)}\mu\tilde{\Psi}(k,q,\mu,x)=S(k,q,\mu,x),
\label{eq:boltzmann_app}
\end{eqnarray}
where
\begin{eqnarray}
S(k,q,\mu,x)
\equiv i\frac{\epsilon(q,x)}{q}\mu\psi(k,x)-\frac{\partial\phi(k,x)}{\partial x}.
\end{eqnarray}
The solution of the first order linear differential equation with the source
term $S(k,q,\mu,x)$ is
\begin{eqnarray}
&&\tilde{\Psi}(k,q,\mu,x)=\tilde{\Psi}(k,q,\mu,x_{i})\exp\left[-i\mu\int^{x}_{x_i}\frac{q}{\epsilon(q,x)}dx'
\right]
+\exp\left[-i\mu\int^{x}_{x_i}\frac{q}{\epsilon(q,x')}dx'\right]
\nonumber\\
&&\times\int^{x}_{x_i}dx'\exp\left[i\mu\int^{x'}_{x_i}\frac{q}{\epsilon(q,x'')}dx''\right]
S(k,q,\mu,x')
\nonumber\\
&=&\tilde{\Psi}(k,q,\mu,x_{i})e^{-i\mu[z(x)-z(x_i)]}
+\int^{x}_{x_i}dx'e^{-i\mu[z(x)-z(x')]}S(k,q,\mu,x'),
\label{eq:boltzmann_soln_app}
\end{eqnarray}
where we define
\begin{eqnarray}
z(x)\equiv\int^x_C\frac{q}{\epsilon(q,x')}dx',
\label{eq:z_app}
\end{eqnarray}
for arbitrary constant, $C\in\mathcal{R}$.

\subsection{massless case with constant $\phi$ and $\psi$}
For massless neutrinos, $\epsilon(q,x)=q$, with constant gravitational potential,
$\dot{\phi}=\dot{\psi}=0$, we can avoid the complexity in the time dependent
source term, $S(q,x)$, and the form of Eq.(\ref{eq:boltzmann_soln_app}) is simplified
to
\begin{eqnarray}
\tilde{\Psi}(k,q,\mu,x)&=&\tilde{\Psi}(k,q,\mu,x_{i})e^{-i\mu(x-x_i)}+
i\mu\psi(k)\int^{x}_{x_i}dx'e^{-i\mu(x-x')}
\nonumber\\
&=&\left[\tilde{\Psi}(k,q,\mu,x_{i})-\psi(k)\right]e^{-i\mu(x-x_i)}+\psi(k).
\label{eq:psi_soln_app}
\end{eqnarray}
Now, we expand Eq.(\ref{eq:psi_soln_app}) by Legendre polynomials, using the
following identities and orthogonality condition,
\begin{eqnarray}
\tilde{\Psi}(k,q,\mu,x)=\sum_l(-i)^l(2l+1)\tilde{\Psi}_l(k,q,x)P_l(\mu),
\label{eq:identity1_app}\\
e^{-i\mu(x-x')}=\sum_l(-i)^l(2l+1)j_l(x-x')P_l(\mu),
\label{eq:identity2_app}\\
\int^{1}_{-1}d\mu P_l(\mu)P_{l'}(\mu)=\frac2{2l+1}\delta_{ll'}.
\label{eq:identity3_app}
\end{eqnarray}
We find
\begin{eqnarray}
\sum_{l'}(-i)^{l'}(2l'+1)\tilde{\Psi}_{l'}(k,q,x)P_{l'}(\mu)&=&
\sum_{l'}\sum_{l''}(-i)^{l'+l''}(2l'+1)(2l''+1)\tilde{\Psi}_{l'}(k,q,x_i)
j_{l''}(x-x_i)P_{l'}(\mu)P_{l''}(\mu)\nonumber\\
&+&\psi(k)\left[1-\sum_{l'}(-i)^{l'}(2l'+1)j_{l'}(x-x_i)P_{l'}(\mu)\right].
\label{eq:psi_exp_app}
\end{eqnarray}
Multiplying the both sides by $P_l(\mu)$ and integrating over $\mu$, we find,
\begin{eqnarray}
\sum_{l}(-i)^{l}\tilde{\Psi}_{l}(k,q,x)&=&
\sum_{l'}\sum_{l''}(-i)^{l'+l''}(2l'+1)(2l''+1)\tilde{\Psi}_{l'}(k,q,x_i)
j_{l''}(x-x_i)
\begin{pmatrix}
l & l' & l''\\
0 & 0 & 0
\end{pmatrix}^2
\nonumber\\
&&+\psi(k)\left[\delta_{l0}-\sum_l(-i)^lj_l(x-x_i)\right],
\nonumber\\
\end{eqnarray}
where we have used the Wigner 3-$j$ symbols to write
\begin{eqnarray}
\int^{1}_{-1}\frac{d\mu}{2}P_{l}(\mu)P_{l'}(\mu)P_{l''}(\mu)=
\begin{pmatrix}
l & l' & l''\\
0 & 0 & 0
\end{pmatrix}^2.
\end{eqnarray}
Therefore, for a given $l$, we have
\begin{eqnarray}
\tilde{\Psi}_{l}(k,q,x)\!\!&=&\!\!
\sum_{l'}\sum_{l''}(-i)^{l'+l''-l}(2l'+1)(2l''+1)\tilde{\Psi}_{l'}(k,q,x_i)
j_{l''}(x-x_i)
\begin{pmatrix}
l & l' & l''\\
0 & 0 & 0
\end{pmatrix}^2
\!\!+\!\psi(k)\left[\delta_{l0}-j_l(x-x_i)\right].
\label{eq:psi_l_massless_app}
\end{eqnarray}
Here, non-zero Wigner 3-$j$ symbols must satisfy the triangular inequalities
such that
\begin{eqnarray}
|l-l'|\le l''\le l+l',
\end{eqnarray}
and from the initial conditions, we have $\tilde{\Psi}_{l'\ge 3}(x_i)=0$.

The exact solutions for $l=0$ and $1$ are
\begin{eqnarray}
\tilde{\Psi}_0(k,q,x)&=&\tilde{\Psi}_0(k,q,x_i)j_0(x-x_i)
-3\tilde{\Psi}_1(k,q,x_i)j_1(x-x_i)
\nonumber\\
&+&5\tilde{\Psi}_2(k,q,x_i)j_2(x-x_i)+\psi(k)\left[1-j_0(x-x_i)\right],\\
\tilde{\Psi}_1(k,q,x)&=&\tilde{\Psi}_0(k,q,x_i)j_1(x-x_i)
+\tilde{\Psi}_1(k,q,x_i)j_0(x-x_i)-2\tilde{\Psi}_2(k,q,x_i)j_1(x-x_i)
\nonumber\\
&-&2\tilde{\Psi}_1(k,q,x_i)j_2(x-x_i)
+3\tilde{\Psi}_2(k,q,x_i)j_3(x-x_i)-\psi(k) j_1(x-x_i).
\end{eqnarray}

\subsection{massive case with constant $\phi$ and $\psi$}
For massive neutrinos, $\epsilon(q,x)\ne q$, with constant gravitational potential,
$\dot{\phi}=\dot{\psi}=0$, we have
\begin{eqnarray}
\sum_{l'}(-i)^{l'}(2l'+1)\tilde{\Psi}_{l'}(k,q,x)P_{l'}(\mu)&=&
\sum_{l'}\sum_{l''}(-i)^{l'+l''}(2l'+1)(2l''+1)\tilde{\Psi}_{l'}(k,q,x_i)
j_{l''}(z-z_i)P_{l'}(\mu)P_{l''}(\mu)\nonumber\\
&+&i\psi(k)\mu\sum_{l'}(-i)^{l'}(2l'+1)P_{l'}(\mu)\int^x_{x_i}dx'
\frac{\epsilon(x')}{q}j_{l'}(z-z'),
\end{eqnarray}
where we have used Eqs.(\ref{eq:identity1_app})$\sim$(\ref{eq:identity3_app}) and
$z-z'=z(x)-z(x')\equiv \int^x_{x'}dx'\frac{q}{\epsilon(q,x')}$ as defined
in Eq.(\ref{eq:z_app}).

Multiplying the both sides by $P_l(\mu)$ and integrating over $\mu$, we find
\begin{eqnarray}
\sum_{l}(-i)^{l}\tilde{\Psi}_{l}(k,q,x)&=&
\sum_{l'}\sum_{l''}(-i)^{l'+l''}(2l'+1)(2l''+1)\tilde{\Psi}_{l'}(k,q,x_i)
j_{l''}(z-z_i)
\begin{pmatrix}
l & l' & l''\\
0 & 0 & 0
\end{pmatrix}^2
\nonumber\\
&-&\psi(k)\sum_l(-i)^l\int^x_{x_i}dx'\frac{\epsilon(q,x')}{q}
\left[
\frac{l}{2l+1}j_{l-1}(z-z')-\frac{l+1}{2l+1}j_{l+1}(z-z')
\right].
\end{eqnarray}
Therefore, for a given $l$, we have
\begin{eqnarray}
\tilde{\Psi}_{l}(k,q,x)&=&
\sum_{l'}\sum_{l''}(-i)^{l'+l''-l}(2l'+1)(2l''+1)\tilde{\Psi}_{l'}(k,q,x_i)
j_{l''}(z-z_i)
\begin{pmatrix}
l & l' & l''\\
0 & 0 & 0
\end{pmatrix}^2
\nonumber\\
&-&\psi(k)\int^x_{x_i}dx'\frac{\epsilon(q,x')}{q}
\left[
\frac{l}{2l+1}j_{l-1}(z-z')-\frac{l+1}{2l+1}j_{l+1}(z-z')
\right].
\label{eq:psi_l_massive_app}
\end{eqnarray}

The exact solutions for $l=0$ and $1$ are
\begin{eqnarray}
\tilde{\Psi}_0(k,q,x)&=&\tilde{\Psi}_0(k,q,x_i)j_0(z-z_i)
-3\tilde{\Psi}_1(k,q,x_i)j_1(z-z_i)
+5\tilde{\Psi}_2(k,q,x_i)j_2(z-z_i)\nonumber\\
&+&\psi(k)\int^x_{x_i}dx'\frac{\epsilon(q,x')}{q}j_1(z-z'),\\
\tilde{\Psi}_1(k,q,x)&=&\tilde{\Psi}_0(k,q,x_i)j_1(z-z_i)
+\tilde{\Psi}_1(k,q,x_i)j_0(z-z_i)-2\tilde{\Psi}_2(k,q,x_i)j_1(z-z_i)
\nonumber\\
&-&2\tilde{\Psi}_1(k,q,x_i)j_2(z-z_i)
+3\tilde{\Psi}_2(k,q,x_i)j_3(z-z_i)
\nonumber\\
&-&\psi(k)\int^x_{x_i}dx'\frac{\epsilon(q,x')}{q}\left[\frac13j_0(z-z')
-\frac23j_2(z-z')\right].
\end{eqnarray}

\subsection{general case}
For more general cases, where neutrinos are either massless or massive,
and $\dot{\phi}\ne 0$ and $\dot{\psi}\ne 0$.
In this case, we use the full expression of the source term,
$S(k,q,\mu,x)\equiv i\frac{\epsilon(q,x)}{q}\mu\psi(k,x)-\frac{\partial\phi(k,x)}{\partial x}$.

Again, we expand Eq.(\ref{eq:boltzmann_soln_app})
with the series of Legendre polynomials with the time dependent source term,
and find
\begin{eqnarray}
\sum_{l'}(-i)^{l'}(2l'+1)\tilde{\Psi}_{l'}(k,q,x)P_{l'}(\mu)&=&
\sum_{l'}\sum_{l''}(-i)^{l'+l''}(2l'+1)(2l''+1)\tilde{\Psi}_{l'}(k,q,x_i)
j_{l''}(z-z_i)P_{l'}(\mu)P_{l''}(\mu)\nonumber\\
&+&i\mu\sum_{l'}(-i)^{l'}(2l'+1)P_{l'}(\mu)\int^x_{x_i}dx'
\frac{\epsilon(x')}{q}\psi(k,x')j_{l'}(z-z')\nonumber\\
&-&\sum_{l'}(-i)^{l'}(2l'+1)P_{l'}(\mu)\int^x_{x_i}dx'
\frac{\partial\phi(k,x')}{\partial x'}j_{l'}(z-z').
\end{eqnarray}

Multiplying the both sides by $P_l(\mu)$ and integrating over $\mu$, we find
\begin{eqnarray}
\sum_{l}(-i)^{l}\tilde{\Psi}_{l}(k,q,x)&=&
\sum_{l'}\sum_{l''}(-i)^{l'+l''}(2l'+1)(2l''+1)\tilde{\Psi}_{l'}(k,q,x_i)
j_{l''}(z-z_i)
\begin{pmatrix}
l & l' & l''\\
0 & 0 & 0
\end{pmatrix}^2
\nonumber\\
&-&\sum_l(-i)^l\int^x_{x_i}dx'\frac{\epsilon(q,x')}{q}\psi(k,x')
\left[
\frac{l}{2l+1}j_{l-1}(z-z')-\frac{l+1}{2l+1}j_{l+1}(z-z')
\right]\nonumber\\
&-&\sum_l(-i)^l\int^x_{x_i}dx'\frac{\partial\phi(k,x')}{\partial x'}j_l(z-z').
\end{eqnarray}
With the recursion relation,
\begin{eqnarray}
\frac{d}{dx}j_l(x)=\frac{l}{2l+1}j_{l-1}(x)-\frac{l+1}{2l+1}j_{l+1}(x),
\end{eqnarray}
we have
\begin{eqnarray}
\tilde{\Psi}_{l}(k,q,x)&=&
\sum_{l'}\sum_{l''}(-i)^{l'+l''-l}(2l'+1)(2l''+1)\tilde{\Psi}_{l'}(k,q,x_i)
j_{l''}(z-z_i)
\begin{pmatrix}
l & l' & l''\\
0 & 0 & 0
\end{pmatrix}^2
\nonumber\\
&-&\int^x_{x_i}dx'
\left[\frac{\epsilon(q,x')}{q}\psi(k,x')-\frac{q}{\epsilon(q,x')}\phi(k,x')
\right]
\left[
\frac{l}{2l+1}j_{l-1}(z-z')-\frac{l+1}{2l+1}j_{l+1}(z-z')
\right]
\nonumber\\
&+&\phi(k,x_i)j_l(z-z_i)-\phi(k,x)\delta_{l0},
\label{eq:psi_l_general_app}
\end{eqnarray}
for a fixed $l$.
We can easily recover the massless and massive case solutions
Eq.(\ref{eq:psi_l_massless_app}) and (\ref{eq:psi_l_massive_app}) from
Eq.(\ref{eq:psi_l_general_app}) with approximations such as
$\dot{\psi}(k,x)=\dot{\phi}(k,x)=0$ and/or $\epsilon(q,x)=q$.
\end{widetext}
%%%%%%%%%%%%%%%%%%%%%%%%%%%%%%%%%%%%%%%%%%%%%%%%%%%%%%%%%%%%%%%%%%

\end{document}